%% file: ms.tex
\documentclass[preprint]{aastex631}
\usepackage{bm, amsmath, natbib}
\usepackage[caption=false,labelformat=empty]{subfig}

\graphicspath{{./}{figures/}}
\usepackage{rotating}

\def\au{{\rm au}}

\def\yr{{\rm yr}}

\def\mas{{\rm mas}}

\def\min{{\rm min}}
\def\rel{{\rm rel}}

\def\eff{{\rm eff}}

\def\B{{\rm B}}
\def\Sc{{\rm S}}
\def\L{{\rm L}}
\def\E{{\rm E}}
\def\bpi{{\bm\pi}}
\def\bmu{{\bm\mu}}

\def\deg{^\circ}
\def\close{{``close A$-$''}}
\def\wide{{``wide A$-$''}}

\begin{document}

\title{First Resolution of Microlensed Images of a Binary-Lens Event}

\correspondingauthor{Subo Dong}
\email{dongsubo@pku.edu.cn}

\author[0009-0007-5754-6206]{Zexuan Wu}
\affil{Department of Astronomy, School of Physics, Peking University,
	5 Yiheyuan Road, Haidian District, Beijing 100871, People's Republic of China \\}
\affil{Kavli Institute of Astronomy and Astrophysics, Peking University,
	5 Yiheyuan Road, Haidian District, Beijing 100871, People's Republic of China \\}
\author[0000-0002-1027-0990]{Subo Dong}
\affil{Department of Astronomy, School of Physics, Peking University,
	5 Yiheyuan Road, Haidian District, Beijing 100871, People's Republic of China \\}
\affil{Kavli Institute of Astronomy and Astrophysics, Peking University,
	5 Yiheyuan Road, Haidian District, Beijing 100871, People's Republic of China \\}
\affil{National Astronomical Observatories, Chinese Academy of Science, 20A Datun Road, Chaoyang District, Beijing 100101, People's Republic of China\\}
\author[0000-0003-2125-0183]{A.~M\'erand}
\affil{European Southern Observatory, Karl-Schwarzschild-Stra{\ss}e 2, D-85748 Garching, Germany\\}
\author{Christopher S. Kochanek}
\affil{Department of Astronomy, Ohio State University, 140 W. 18th Ave., Columbus, OH 43210, USA\\}
\affil{Center for Cosmology and AstroParticle Physics (CCAPP), The Ohio State University, 191 W. Woodruff Avenue, Columbus, OH 43210, USA.\\}
\author[0000-0001-7016-1692]{Przemek Mr{\'o}z}
\affil{Astronomical Observatory, University of Warsaw, Al. Ujazdowskie 4, 00-478 Warszawa, Poland\\}
\author{Jinyi Shangguan}
\affil{Max Planck Institute for Extraterrestrial Physics, Giessenbachstra{\ss}e 1, D-85748 Garching, Germany\\}
\author{Grant Christie}
\affil{Auckland Observatory, Auckland, New Zealand\\}
\author[0000-0001-5603-6895]{Thiam-Guan Tan}
\affil{Perth Exoplanet Survey Telescope, Perth, Australia\\}
\author[0000-0003-3978-1409]{Thomas Bensby}
\affil{Lund Observatory, Division of Astrophysics, Department of Physics, Lund University, Box 118, SE-22100 Lund, Sweden\\}
\author[0000-0001-7516-4016]{Joss Bland-Hawthorn}
\affil{Sydney Institute for Astronomy, School of Physics, A28, The University of Sydney, NSW 2006, Australia\\}
\affil{ARC Centre of Excellence for All Sky Astrophysics in 3 Dimensions (ASTRO 3D), Australia\\}
\author[0000-0002-4031-8553]{Sven Buder}
\affil{Research School of Astronomy and Astrophysics, The Australian National University, Canberra ACT2611, Australia\\}
\affil{ARC Centre of Excellence for All Sky Astrophysics in 3 Dimensions (ASTRO 3D), Australia\\}
\author{Frank Eisenhauer}
\affil{Max Planck Institute for Extraterrestrial Physics, Giessenbachstra{\ss}e 1, D-85748 Garching, Germany\\}
\affil{Department of Physics, Technical University of Munich, 85748 Garching, Germany\\}
\author{Andrew P. Gould}
\affil{Max Planck Institute for Astronomy, K\"{o}nigstuhl 17, 69117 Heidelberg, Germany\\}
\affil{Department of Astronomy, Ohio State University, 140 W. 18th Ave., Columbus, OH 43210, USA\\}
\author[0000-0003-2533-6056]{Janez Kos}
\affil{Faculty of Mathematics and Physics, University of Ljubljana, Jadranska 19, 1000 Ljubljana, Slovenia\\}
\author{Tim Natusch}
\affil{Mathematical Sciences Department, Auckland University of Technology, Auckland, New Zealand\\}
\affil{Auckland Observatory, Auckland, New Zealand\\}
\author[0000-0002-0920-809X]{Sanjib Sharma}
\affil{Sydney Institute for Astronomy, School of Physics, A28, The University of Sydney, NSW 2006, Australia\\}
\affil{ARC Centre of Excellence for All Sky Astrophysics in 3 Dimensions (ASTRO 3D), Australia\\}
\author[0000-0001-5207-5619]{Andrzej Udalski}
\affil{Astronomical Observatory, University of Warsaw, Al. Ujazdowskie 4, 00-478 Warszawa, Poland\\}
\author{J. Woillez}
\affil{European Southern Observatory, Karl-Schwarzschild-Stra{\ss}e 2, D-85748 Garching, Germany\\}
\author{David A. H. Buckley}
\affil{South African Astronomical Observatory, P.O. Box 9, Observatory 7935, Cape Town, South Africa\\}
\author{I. B. Thompson}
\affil{Carnegie Observatories, 813 Santa Barbara Street, Pasadena, CA 91101-1292, USA}
\nocollaboration{20}
\author{Karim Abd El Dayem}
\affil{LESIA, Observatoire de Paris, Universit\'{e} PSL, Sorbonne Universit\'{e}, Universit\'{e} Paris Cit\'{e}, CNRS, 5 place Jules Janssen, 92195 Meudon, France\\}

\author{Anthony Berdeu}
\affil{LESIA, Observatoire de Paris, Universit\'{e} PSL, Sorbonne Universit\'{e}, Universit\'{e} Paris Cit\'{e}, CNRS, 5 place Jules Janssen, 92195 Meudon, France\\}

\author{Jean-Philippe Berger}
\affil{Univ. Grenoble Alpes, CNRS, IPAG, 38000 Grenoble, France\\}

\author{Guillaume Bourdarot}
\affil{Max Planck Institute for Extraterrestrial Physics, Giessenbachstra{\ss}e 1, D-85748 Garching, Germany\\}

\author{Wolfgang Brandner}
\affil{Max Planck Institute for Astronomy, K\"{o}nigstuhl 17, 69117 Heidelberg, Germany\\}

\author{Richard I. Davies}
\affil{Max Planck Institute for Extraterrestrial Physics, Giessenbachstra{\ss}e 1, D-85748 Garching, Germany\\}

\author{Denis Defr\`{e}re}
\affil{Institute of Astronomy, KU Leuven, Celestijnenlaan 200D, 3001, Leuven, Belgium\\}

\author{Catherine Dougados}
\affil{Univ. Grenoble Alpes, CNRS, IPAG, 38000 Grenoble, France\\}

\author{Antonia Drescher}
\affil{Max Planck Institute for Extraterrestrial Physics, Giessenbachstra{\ss}e 1, D-85748 Garching, Germany\\}

\author{Andreas Eckart}
\affil{1st Institute of Physics, University of Cologne, Z\"{u}lpicher Stra{\ss}e 77, 50937 Cologne, Germany\\}
\affil{Max Planck Institute for Radio Astronomy, Auf dem H\"{u}gel 69, 53121 Bonn, Germany\\}

\author{Maximilian Fabricius}
\affil{Max Planck Institute for Extraterrestrial Physics, Giessenbachstra{\ss}e 1, D-85748 Garching, Germany\\}

\author{Helmut Feuchtgruber}
\affil{Max Planck Institute for Extraterrestrial Physics, Giessenbachstra{\ss}e 1, D-85748 Garching, Germany\\}

\author{Natascha M. F\"{o}rster Schreiber}
\affil{Max Planck Institute for Extraterrestrial Physics, Giessenbachstra{\ss}e 1, D-85748 Garching, Germany\\}

\author{Paulo Garcia}
\affil{Faculdade de Engenharia, Universidade do Porto, R Dr. Roberto Frias s/n, 4200-465 Porto, Portugal\\}
\affil{CENTRA – Centro de Astrof\'{i}sica e Gravita\c{c}\~{a}o, IST, Universidade de Lisboa, 1049-001 Lisboa, Portugal \\}

\author{Reinhard Genzel}
\affil{Max Planck Institute for Extraterrestrial Physics, Giessenbachstra{\ss}e 1, D-85748 Garching, Germany\\}
\affil{Departments of Physics and Astronomy, Le Conte Hall, University of California, Berkeley, CA 94720, USA\\}

\author{Stefan Gillessen}
\affil{Max Planck Institute for Extraterrestrial Physics, Giessenbachstra{\ss}e 1, D-85748 Garching, Germany\\}

\author{Gernot Hei{\rm \ss}el}
\affil{LESIA, Observatoire de Paris, Universit\'{e} PSL, Sorbonne Universit\'{e}, Universit\'{e} Paris Cit\'{e}, CNRS, 5 place Jules Janssen, 92195 Meudon, France\\}

\author{Sebastian H\"{o}nig}
\affil{School of Physics \& Astronomy, University of Southampton, University Road, Southampton SO17 1BJ, UK\\}

\author{Mathis Houlle}
\affil{Universit\'{e} C\`{o}te d’Azur, Observatoire de la C\`{o}te d’Azur, CNRS, Laboratoire Lagrange, CS 34229, F-06304 Nice cedex 4, France\\}

\author{Pierre Kervella}
\affil{LESIA, Observatoire de Paris, Universit\'{e} PSL, Sorbonne Universit\'{e}, Universit\'{e} Paris Cit\'{e}, CNRS, 5 place Jules Janssen, 92195 Meudon, France\\}

\author{Laura Kreidberg}
\affil{Max Planck Institute for Astronomy, K\"{o}nigstuhl 17, 69117 Heidelberg, Germany\\}

\author{Sylvestre Lacour}
\affil{LESIA, Observatoire de Paris, Universit\'{e} PSL, Sorbonne Universit\'{e}, Universit\'{e} Paris Cit\'{e}, CNRS, 5 place Jules Janssen, 92195 Meudon, France\\}
\affil{European Southern Observatory, Karl-Schwarzschild-Stra{\ss}e 2, D-85748 Garching, Germany\\}

\author{Olivier Lai}
\affil{Universit\'{e} C\`{o}te d’Azur, Observatoire de la C\`{o}te d’Azur, CNRS, Laboratoire Lagrange, CS 34229, F-06304 Nice cedex 4, France\\}

\author{Romain Laugier}
\affil{Institute of Astronomy, KU Leuven, Celestijnenlaan 200D, 3001, Leuven, Belgium\\}

\author{Jean-Baptiste Le Bouquin}
\affil{Univ. Grenoble Alpes, CNRS, IPAG, 38000 Grenoble, France\\}

\author{James Leftley}
\affil{Universit\'{e} C\`{o}te d’Azur, Observatoire de la C\`{o}te d’Azur, CNRS, Laboratoire Lagrange, CS 34229, F-06304 Nice cedex 4, France\\}

\author{Bruno Lopez}
\affil{Universit\'{e} C\`{o}te d’Azur, Observatoire de la C\`{o}te d’Azur, CNRS, Laboratoire Lagrange, CS 34229, F-06304 Nice cedex 4, France\\}

\author{Dieter Lutz}
\affil{Max Planck Institute for Extraterrestrial Physics, Giessenbachstra{\ss}e 1, D-85748 Garching, Germany\\}

\author{Felix Mang}
\affil{Max Planck Institute for Extraterrestrial Physics, Giessenbachstra{\ss}e 1, D-85748 Garching, Germany\\}

\author{Florentin Millour}
\affil{Universit\'{e} C\`{o}te d’Azur, Observatoire de la C\`{o}te d’Azur, CNRS, Laboratoire Lagrange, CS 34229, F-06304 Nice cedex 4, France\\}

\author{Miguel Montarg\`{e}s}
\affil{LESIA, Observatoire de Paris, Universit\'{e} PSL, Sorbonne Universit\'{e}, Universit\'{e} Paris Cit\'{e}, CNRS, 5 place Jules Janssen, 92195 Meudon, France\\}

\author{Hugo Nowacki}
\affil{Univ. Grenoble Alpes, CNRS, IPAG, 38000 Grenoble, France\\}

\author{Mathias Nowak}
\affil{LESIA, Observatoire de Paris, Universit\'{e} PSL, Sorbonne Universit\'{e}, Universit\'{e} Paris Cit\'{e}, CNRS, 5 place Jules Janssen, 92195 Meudon, France\\}

\author{Thomas Ott}
\affil{Max Planck Institute for Extraterrestrial Physics, Giessenbachstra{\ss}e 1, D-85748 Garching, Germany\\}

\author{Thibaut Paumard}
\affil{LESIA, Observatoire de Paris, Universit\'{e} PSL, Sorbonne Universit\'{e}, Universit\'{e} Paris Cit\'{e}, CNRS, 5 place Jules Janssen, 92195 Meudon, France\\}

\author{Karine Perraut}
\affil{Univ. Grenoble Alpes, CNRS, IPAG, 38000 Grenoble, France\\}

\author{Guy Perrin}
\affil{LESIA, Observatoire de Paris, Universit\'{e} PSL, Sorbonne Universit\'{e}, Universit\'{e} Paris Cit\'{e}, CNRS, 5 place Jules Janssen, 92195 Meudon, France\\}

\author{Romain Petrov}
\affil{Universit\'{e} C\`{o}te d’Azur, Observatoire de la C\`{o}te d’Azur, CNRS, Laboratoire Lagrange, CS 34229, F-06304 Nice cedex 4, France\\}

\author{Pierre-Olivier  Petrucci}
\affil{Univ. Grenoble Alpes, CNRS, IPAG, 38000 Grenoble, France\\}

\author{Nicolas Pourre}
\affil{Univ. Grenoble Alpes, CNRS, IPAG, 38000 Grenoble, France\\}

\author{Sebastian Rabien}
\affil{Max Planck Institute for Extraterrestrial Physics, Giessenbachstra{\ss}e 1, D-85748 Garching, Germany\\}

\author{Diogo C. Ribeiro}
\affil{Max Planck Institute for Extraterrestrial Physics, Giessenbachstra{\ss}e 1, D-85748 Garching, Germany\\}

\author{Sylvie Robbe-Dubois}
\affil{Universit\'{e} C\`{o}te d’Azur, Observatoire de la C\`{o}te d’Azur, CNRS, Laboratoire Lagrange, CS 34229, F-06304 Nice cedex 4, France\\}

\author{Matteo  Sadun Bordoni}
\affil{Max Planck Institute for Extraterrestrial Physics, Giessenbachstra{\ss}e 1, D-85748 Garching, Germany\\}

\author{Daryl Santos}
\affil{Max Planck Institute for Extraterrestrial Physics, Giessenbachstra{\ss}e 1, D-85748 Garching, Germany\\}

\author{Jonas Sauter}
\affil{Max Planck Institute for Astronomy, K\"{o}nigstuhl 17, 69117 Heidelberg, Germany\\}

\author{Jules Scigliuto}
\affil{Universit\'{e} C\`{o}te d’Azur, Observatoire de la C\`{o}te d’Azur, CNRS, Laboratoire Lagrange, CS 34229, F-06304 Nice cedex 4, France\\}

\author{Taro T. Shimizu}
\affil{Max Planck Institute for Extraterrestrial Physics, Giessenbachstra{\ss}e 1, D-85748 Garching, Germany\\}

\author{Christian Straubmeier}
\affil{1st Institute of Physics, University of Cologne, Z\"{u}lpicher Stra{\ss}e 77, 50937 Cologne, Germany\\}

\author{Eckhard Sturm}
\affil{Max Planck Institute for Extraterrestrial Physics, Giessenbachstra{\ss}e 1, D-85748 Garching, Germany\\}

\author{Matthias Subroweit}
\affil{1st Institute of Physics, University of Cologne, Z\"{u}lpicher Stra{\ss}e 77, 50937 Cologne, Germany\\}

\author{Calvin Sykes}
\affil{School of Physics \& Astronomy, University of Southampton, University Road, Southampton SO17 1BJ, UK\\}

\author{Linda Tacconi}
\affil{Max Planck Institute for Extraterrestrial Physics, Giessenbachstra{\ss}e 1, D-85748 Garching, Germany\\}

\author{Fr\'{e}d\'{e}ric Vincent}
\affil{LESIA, Observatoire de Paris, Universit\'{e} PSL, Sorbonne Universit\'{e}, Universit\'{e} Paris Cit\'{e}, CNRS, 5 place Jules Janssen, 92195 Meudon, France\\}

\author{Felix Widmann}
\affil{Max Planck Institute for Extraterrestrial Physics, Giessenbachstra{\ss}e 1, D-85748 Garching, Germany\\}

\collaboration{58}{(the GRAVITY+ collaboration)}

\begin{abstract}
	We resolve the multiple images of the binary-lens microlensing event ASASSN-22av using the GRAVITY instrument of the Very Large Telescope Interferometer (VLTI). The light curves show weak binary-lens perturbations, complicating the analysis, but the joint modeling with the VLTI data breaks several degeneracies, arriving at a strongly favored solution. Thanks to precise measurements of angular Einstein radius $\theta_\E = 0.724\pm0.002$\,mas and microlens parallax, we determine that the lens system consists of two M dwarfs with masses of $M_1= 0.258\pm{0.008}\, M_\odot$ and $M_2 = 0.130\pm{0.007}\, M_\odot$, a projected separation of $r_\perp = 6.83\pm0.31\,\au$ and a distance of $D_\L=2.29\pm0.08$\,kpc. The successful VLTI observations of ASASSN-22av open up a new path for studying intermediate-separation (i.e., a few astronomical units) stellar-mass binaries, including those containing dark compact objects such as neutron stars and stellar-mass black holes.
\end{abstract}


\section{Introduction} \label{sec:intro}
In recent years, the enhanced sensitivity of advanced instrumentation \citep{Eisenhauer2023} at the Very Large Telescope Interferometer (VLTI) of the European Southern Observatory (ESO) has facilitated the interferometric resolution of microlensed images \citep{vltimicrolens1}, with the first successful case achieved by \citet{Dong2019} using the GRAVITY instrument \citep{gravity2017}. By enabling direct mass determinations, interferometric microlensing opens up a new path into studying Galactic stellar-mass systems, including dark compact objects such as neutron stars and black holes.

The observable effect of microlensing \citep{Einstein1936, Paczynski1986} is the magnified flux of a background star (i.e., the ``source'') due to the gravitational deflection by a foreground object (i.e., the ``lens''), and thus microlensing is a sensitive probe of the mass of the lens system, irrespective of its brightness. However, the most easily extracted quantity, the Einstein timescale $t_\E$ is a degenerate combination of the lens mass $M_\L$, the lens-source relative parallax $\pi_\rel$ and the proper motion $\mu_\rel$:
\begin{equation}
	\label{eq:tE_thetaE}
	t_\E=\frac{\theta_\E}{\mu_\rel};
	\quad
	\theta_\E\equiv \sqrt{\kappa M_\L\pi_\rel};
	\quad
	\kappa\equiv \frac{4 G}{c^2\au}\simeq 8.144 \frac{\mas}{M_\odot},
\end{equation}
where $\theta_\E$ is the angular Einstein radius, which is of the order of milliarcseconds for stellar-mass microlenses in the Galaxy.
To break this degeneracy, a common approach is to measure both $\theta_\E$ and the so-called ``microlens parallax'' $\bpi_\E$ to determine the lens mass \citep{Gould2000, Gould2004}:
\begin{equation}
	\label{eq:M_L}
	M_\L=\frac{\theta_\E}{\kappa\pi_\E},
\end{equation}
where $\bpi_\E$ is a two-dimensional vector \citep{Gould2004} along the same direction as the lens-source relative proper motion vector $\bmu_\rel$:
\begin{equation}
	\label{eq:piE}
	\bpi_\E \equiv \frac{\pi_\rel}{\theta_\E}\frac{\bmu_\rel}{\mu_\rel}.
\end{equation}
The key advantage of the interferometric microlensing method is that it yields not only precise $\theta_\E$, from measuring the angular separation of images, but also constraints on the direction of $\bpi_\E$.

One often obtains ``one-dimensional'' microlens parallax from the light curves; that is, the component of $\bpi_\E$ parallel to the Earth's acceleration is generally much better constrained than the perpendicular component \citep{Gould1d1994, Smith2003, Gould2004}. As demonstrated by \citet{Dong2019}, \citet{Zang2020}, and \citet{Cassan2022}, VLTI microlensing can drastically improve the $\bpi_\E$ measurements from accurate directional constraints via measuring the position angles of images.

A long-adopted method for determining $\theta_\E$ is via the ``finite-source effects'' \citep{Gould1994a}, when the source transits/approaches a caustic (a set of positions that, for a point source, induce infinite magnification).  Modeling the finite-source effects in the microlensing light curve measures the ratio between $\theta_\E$ and the angular radius of the source star $\theta_*$, where $\theta_*$ can be derived from the photometric properties of the source.  For a single lens, the caustic is  point-like, so finite-source detections in stellar single-lens events are usually restricted to rare cases reaching a high peak magnification of $A_{\rm peak}\gtrsim100$. In contrast, binary lenses have extended caustic structures, and thus finite-source effects manifest more often in binary-lens events. The first microlens mass measurement by \citet{An2002} is from such a caustic-crossing binary-lens event. However, this method is ineffective for studying events with weak binary-lens perturbations (e.g., non-caustic-crossing). One such weak-binary event was OGLE-2005-SMC-001, for which \citet{Dong2007} detected the space-based microlens parallax \citep{Refsdal1966, Gould1994b} for the first time, and since no finite-source effects were detected, it was not possible to definitively determine the lens mass due to the absence of a $\theta_\E$ measurement. While the exact fraction of weak binary-lens events is still an open question, \citet{Zhu2014} predicted and \citet{Jung2022} confirmed that half of planetary events would lack caustic crossings, despite their continuous and high-cadence coverages, implying that weak binaries may encompass a significant fraction of binary-lens parameter space.

With milliarcsecond resolution, VLTI can resolve the microlensed images with angular separations around $\gtrsim2\theta_\E$ for a stellar-mass-lens microlensing event \citep{vltimicrolens1,vltimicrolens2,vltimicrolens3,vltimicrolens4} and then can determine $\theta_\E$ at $\sim$percent precision. \citet{Dong2019} measured $\theta_\E = 1.87 \pm 0.03\,\mas$ for TCP J05074264+2447555 (aka Kojima-1) with VLTI GRAVITY, leading to a mass measurement of $M_\L = 0.495 \pm 0.063\,M_\sun$ when combined with the microlens parallax \citep{Zang2020}.
\citet{Cassan2022} obtained $\theta_\E = 0.7650 \pm 0.0038\,\mas$ for Gaia19bld \citep{Rybicki2022} with VLTI PIONIER \citep{LeBouquin2011} and then determined that  $M_\L = 1.147 \pm 0.029\,M_\sun$. Both events were exceptionally bright, since VLTI observations were limited by the instrument's sensitivity, with $K\lesssim 10.5$ for the on-axis mode of GRAVITY and  $H\lesssim7.5$ for PIONIER. GRAVITY has recently been upgraded to enable dual-beam capability (known as ``GRAVITY Wide''; \citealt{gravitywide}) with a boosted sensitivity, and using GRAVITY wide, \citet{Mroz24} measured that OGLE-2023-BLG-0061/KMT-2023-BLG-0496 has $\theta_\E = 1.280\pm0.009$\,mas and inferred its lens mass $M_\L=0.472\pm0.012\,{\rm M_\odot}$. Future upgrades to GRAVITY+ are expected to significantly improve the sensitivity \citep{gravityplus} and consequently the number of accessible microlensing events.

The VLTI microlensing observations published so far have been exclusively on single-lens events (or hosts of extremely low-mass companions). Here, we analyze ASASSN-22av, the {\it first binary-lens} microlensing event with multiple images resolved by VLTI.

\medskip
\section{Observations and Data Reduction} \label{sec:obs}

ASASSN-22av was discovered as a microlensing event candidate\footnote{\url{https://www.astronomy.ohio-state.edu/asassn/transients.html}} on UT 2022-01-19.10 by the All-Sky Automatic Survey for Supernovae \citep[ASAS-SN;][]{Shappee2014}. Its Gaia DR3 \citep{GaiaDR3} \texttt{source\_id} is 5887701839850363776 and coordinates are $(\alpha ,\delta)_{\rm J2000} = (15^{\rm h} 01^{\rm m} 00.66^{\rm s}, -54\arcdeg 23\arcmin 59.96\arcsec)$, corresponding to Galactic coordinates $(l, b)_{\rm J2000} = (321.\deg14522, 3.\deg82467)$. Following its discovery, we modeled the real-time ASAS-SN light curve\footnote{\url{https://asas-sn.osu.edu}} \citep{asassn_v1} and found that it was consistent with a point-source point-lens (PSPL; \citealt{Paczynski1986}) model, reaching a peak magnification of $A_{\rm peak}\sim20$--$30$ around UT 2022 January 20. Its baseline is at $K_s=9.27\pm0.02$ (Two Micron All Sky Survey or 2MASS ID: J15010066$-$5423599) and $G=13.82$, and the magnified brightness was well above the limit for conducting on-axis GRAVITY observations with Multi-Application Curvature Adaptive Optics (MACAO).

We began to conduct VLTI GRAVITY observations
on the night of UT 2022 January 21 and subsequently made ongoing observations until 2022 January 25. We discuss the VLTI observations and data reduction in  \S~\ref{sec:vltidata}. We also started photometric follow-up observations on UT 2022 January 20 (see \S~\ref{sec:photometry} for detailed descriptions of photometric data). On UT 2022 January 21, our light-curve modeling suggested significant deviations from the best-fit PSPL model, and we found that a finite-source point-lens (FSPL) model was now preferred. Aiming at performing ``stellar atmosphere tomography''  \citep[see, e.g.,][and references therein]{Albrow2001, Afonso2001} near the caustic exit, we initiated a campaign of time-series high-resolution spectroscopy.  On UT 2022 January 24, we noticed that the light curve declined faster than expected from the best-fit FSPL model and considered the possibility of a binary-lens perturbation, which was confirmed by our later analysis (see \S~\ref{sec:lc_analysis} for detailed discussions on light-curve modeling). We obtained high-resolution spectra from the High Resolution Spectrograph (HRS) on the Southern African Large Telescope (SALT; \citealt{SALT}),  the MIKE spectrograph \citep{MIKE} on the Magellan Clay 6.5~m telescope, the High Efficiency and Resolution Multi Element Spectrograph (HERMES; \citealt{HERMES})  on the Anglo-Australian Telescope (AAT), and the HARPS spectrograph \citep{HARPS} on the ESO's 3.6~m telescope, from UT 2022 January 25 to UT 2022 January 30. The full description of the spectroscopic data and the stellar atmosphere tomography analysis will be presented in a future paper. In this work, we utilize the stellar parameters of the source extracted from the HERMES data presented in \S~\ref{sec:spectrum}.

\subsection{Interferometric Data}
\label{sec:vltidata}
We observed ASASSN-22av with the GRAVITY instrument \citep{gravity2017} and four 8 m Unit Telescopes (UTs) on the four nights of 2022 January 21, 22, 24, and 25. The observations were executed with the single-field on-axis mode at medium resolution ($R=\lambda / \Delta \lambda \simeq 500)$. We use all data except those on 2022 January 21, when there was significant flux loss in UT2 and poor fringe-tracking performance. The remaining three nights of data are reduced with the standard GRAVITY pipeline (version 1.6.6), and there are two, four, and two exposures on the nights of 22, 24, and 25, respectively. Throughout this paper, we refer to these three nights as Nights 1, 2, and 3, respectively. We first use the Python script \texttt{run\_gravi\_reduced.py} to reduce the raw data up to the application of the pixel-to-visibility matrix (P2VM). The default options are used except for adopting \texttt{--gravity\_vis.output-phase-sc=SELF\_VISPHI} to calculate the internal differential phase between each spectral channel. The pipeline performed the bias and sky subtraction, flat-fielding, wavelength calibration, and spectral extraction. Application of the P2VM converts the pixel detector counts into complex visibilities, taking into account all instrumental effects, including relative throughput, coherence, phase shift, and crosstalk.
The dark, bad-pixel, flat-field, wavelength calibration, and P2VM matrix data are reduced from the daily calibration data obtained close in time to our observations.
We then use \texttt{run\_gravi\_trend.py} to calibrate the closure phase with the calibrator observation of HD 134122 next to the science target observations.

The observations employed four UTs and thus there are four closure phases for every spectral channel.
In principle, the four closure phases are not independent and should sum up to zero.
We examine the calibrator data and find that the standard deviation of the sum of the closure phases is $\sim 0.\deg15$, comparable to the pipeline-reported closure-phase uncertainties in the science data.
Thus, we take the simplified approach and treat the four closure phases as uncorrelated observables rather than the more sophisticated approach of treating the correlated noise employed by \citet{Kammerer2020}.
On each night, we have two exposures for the calibrator star, and we find the closure-phase differences of the two exposures have scatters of $0.\deg21, 0.\deg25, 0.\deg35$ on Night 1, 2, and 3, respectively. We adopt these values as the uncertainties introduced in the calibration process and add them quadratically to the closure-phase errors.

The medium-resolution spectrograph samples 233 wavelengths spanning $2.0-2.4\,\rm \mu m$.
To speed up computation, the 233 closure phases are binned into 10 phases in our modeling process. The first and last phases are discarded due to the low signal-to-noise ratios and significant deviations from the model.
See Appendix~\ref{appendix:binning} for the description of the binning process.

\subsection{Photometric Data}
\label{sec:photometry}
We obtained follow-up imaging data taken with the 1 m telescopes of the Las Cumbres Observatory Global Telescope Network \citep[LCOGT;][]{Brown2013} in the Sloan $g'r'i'$ filters, the 0.4 m telescope of Auckland Observatory (AO) at Auckland (New Zealand) in the $g'r'i'$ filters, and the 30 cm Perth Exoplanet Survey Telescope (PEST) in the $g'r'i'$ filters.

We use the \texttt{PmPyeasy} pipeline \citep{Chen2022} to do the photometry on these follow-up images.
For LCOGT and PEST data, we conduct image subtractions with the \texttt{hotpants}\footnote{\url{https://github.com/acbecker/hotpants}} code \citep{Becker2015}, and then we employ aperture photometry on the subtracted images.
For the AO data, we performed point-spread function (PSF) photometry using \texttt{DoPHOT} \citep{Schechter1993}.

We coadded and analyzed the $g'$-band ASAS-SN images using the standard ASAS-SN photometric pipeline based on the ISIS image-subtraction code \citep{alard_luption1998, alard_image_sub2000}. The baseline ($g' \sim15.5$) is too faint to obtain precise photometry for ASAS-SN, and we analyze the magnified ASAS-SN epochs with ${\rm HJD'= HJD} - 2450000\in (9577.857, 9677.608)$. There are data from three ASAS-SN cameras ({\tt be}, {\tt bi} and {\tt bm}), and they are analyzed independently in the modeling.

This event was announced\footnote{\url{https://gsaweb.ast.cam.ac.uk/alerts/alert/Gaia22ahy/}} as Gaia22ahy by {Gaia} Science Alerts \citep[GSA;][]{Hodgkin2021} on UT 2022-01-25.4.
The {Gaia} $G$-band light curve available on GSA website does not have uncertainties, and we adopt the same error bar ($\sigma_{G} = 0.02$) for all data points. We noticed that the data in 2015 exhibit systematically larger scatter than the rest of the {Gaia} baseline. By inspecting a few other GSA microlensing light curves \citep{Gaiamicrolensing}, we found similar patterns at the beginning of the {Gaia} mission, possibly due to the continued contamination by water ice \citep{Gaia2016}. We conservatively remove all {Gaia} data before the last decontamination run on 2016 August 22 \citep{Gaiaphotometry} from our analysis.

We follow the standard procedure \citep{Yee2012} of renormalizing the error bars such that $\chi^2/$degrees of freedom (dof) is unity for each data set for the best-fit model. We reject the 3$\sigma$ outliers and thereby exclude four out of 983 photometric data points from the analysis (the rejected data consist of one data point each from LCOGT $r'$ and $i'$ and one epoch each from the ASAS-SN {\tt bi} and {\tt bm} cameras). Figure~\ref{fig:lc} shows the multiband light curves of ASASSN-22av.

\subsection{HERMES Spectroscopic Data and Stellar Parameters}
\label{sec:spectrum}

We obtained a high-resolution optical spectrum (\texttt{sobject\_id} 220125001601051) on UT 2022-01-25.75 as part of ongoing observations of the Galactic Archaeology with HERMES survey \citep{DeSilva2015}. The total exposure time is 3600\,s. The HERMES spectrograph on the 3.9 m AAT at Siding Spring Observatory covers four wavelength regions (4713--4903, 5648--5873, 6478--6737, and 7585--7887 \AA) that can show absorption features from up to 31 elements, including the two strong Balmer lines $\mathrm{H\alpha}$ and $\mathrm{H\beta}$.

We use the observations and analysis of the most recent analysis cycle, which will be published as the fourth GALAH data release (\citealt{Buder2024}). The spectra are reduced using a similar analysis as that for the third data release \citep{Buder2021}, with a custom-built reduction pipeline \citep{Kos2017}. Stellar parameters---that is, effective temperature ($T_\mathrm{eff}$), surface gravity ($\log g$), and radial, microturbulence, and rotational velocities ($v_\mathrm{rad}$, $v_\mathrm{mic}$ and $v \sin i$), as well as up to 31 elemental abundances---are then simultaneously estimated by minimizing the weighted residuals between the observed and synthetic stellar spectra. The latter are interpolated with a high-dimensional neural network \citep[compare to][]{Ting2019}, trained on synthetic stellar spectra generated with the spectrum synthesis code Spectroscopy Made Easy \citep{Piskunov2017}, and degraded to the measured instrumental resolution of the spectrum. This change with respect to the third data release, which used wavelength regions of carefully selected unblended lines, allows 94\% of the observed wavelength range to be used and thus considerably increases the measurement precision.

The spectroscopic fit also includes photometric and astrometric information from the 2MASS \citep{Skrutskie2006} and {Gaia} surveys \citep{Lindegren2021a} to constrain the surface gravity. This leads to more accurate stellar parameters, as shown in the previous HERMES analyses \citep{Buder2018, Buder2019, Buder2021}, particularly for cool giant stars such as the target of this study, where molecular features complicate the spectroscopic analysis. The fit is very good, with the signal-to-noise ratio increasing across the four wavelength regions (16, 107, 194, and 254), as can be seen from Figure~\ref{fig:hermes_spectrum}. We do not expect a good fit to the Balmer-line cores or the lithium lines at $4861$, $6563$, and $6708\,\text{\AA}$, respectively. No quality concerns are raised by the automatic pipeline (\texttt{flag\_sp} = 0).

The final radial velocity of $v_\mathrm{rad} = -72.44 \pm 0.18\,\mathrm{km\,s^{-1}}$ is in very good agreement with the value from the {Gaia} DR3 spectrum---that is, $v_{\mathrm{rad, Gaia}} = -71.8 \pm 1.0\,\mathrm{km\,s^{-1}}$ \citep{RecioBlanco2023}. The final stellar parameters are $T_\mathrm{eff} = 3776 \pm 67\,\mathrm{K}$, $\log g = 1.05 \pm 0.12$, $\mathrm{[Fe/H]} = -0.148 \pm 0.053$, $v_\mathrm{mic} = 1.82 \pm 0.28 \,\mathrm{km\,s^{-1}}$, and $v \sin i = 5.5 \pm 1.4 \,\mathrm{km\,s^{-1}}$. The stellar abundances are close to the solar values; for example, $\mathrm{[Mg/Fe]} = 0.007 \pm 0.011$.

%%%%%%%%%%%%%%%%%%%%%%%%%%%%%%%%%%%%%%%%%%%%%%%
\bigskip
\section{Light-curve Analysis}
\label{sec:lc_analysis}
\subsection{Single-lens Model}
\label{subsec:espl_analysis}

In a microlensing event, the flux is modeled as
\begin{equation}
	f(t) = f_\Sc \cdot A(t) + f_\B,
\end{equation}
where $A(t)$ is the magnification as a function of time, and $f_\Sc$ and $f_\B$ are the source flux and the blended flux within the PSF, respectively.
In the simplest single-lens single-source (1L1S) model, i.e., the PSPL model \citep{Paczynski1986}, the magnification is given by
\begin{equation}
	A_{\rm PSPL} = \frac{u^2 + 2}{u\sqrt{u^2 + 4}}; \qquad u(t) = \sqrt{u_0^2 + \frac{(t-t_0)^2}{t_\E^2}},
\end{equation}
where $u$ is the angular lens-source separation normalized by $\theta_\E$, and ($t_0$, $u_0$) are the time of the closest source-lens approach and the impact parameter normalized by $\theta_\E$, respectively.

\begin{figure}[htbp]
	\centering
	\includegraphics[width=0.8\textwidth]{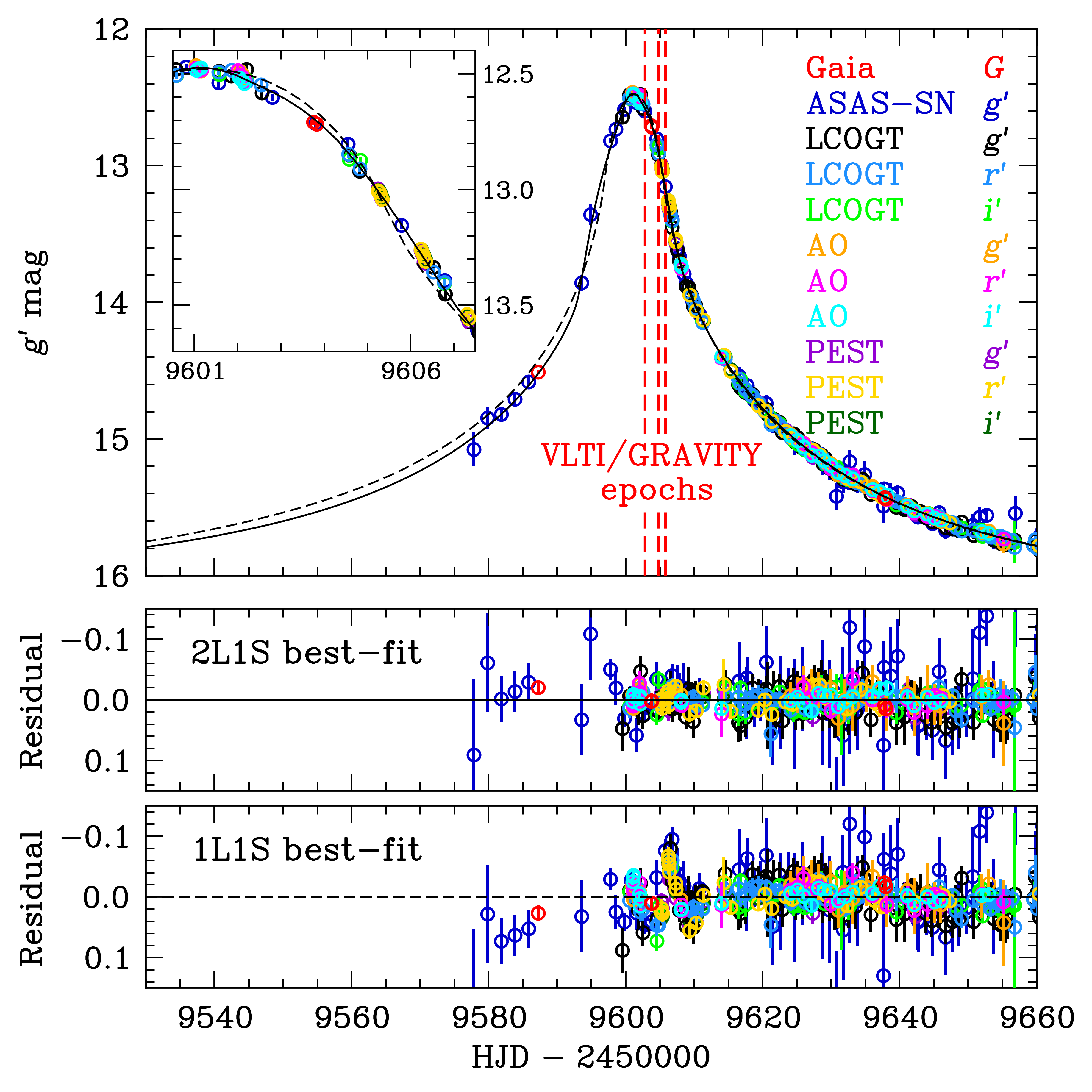}
	\caption{Top: the multiband light curves of the microlensing event ASASSN-22av, which are well described by the best-fit 2L1S model (black solid line) from the joint VLTI and light-curve analysis. The light curves deviate from the best-fit 1L1S model with finite-source effects (black dashed line) near the peak, as shown with the enlarged view in the inset. We analyze three nights of VLTI GRAVITY observations taken after the peak, and their epochs are marked with the red dashed lines.
		Middle: the residuals of the best-fit  2L1S model.
		Bottom: the best-fit 1L1S model shows significant residuals for the $\sim 10$ days around the peak.}
	\label{fig:lc}
\end{figure}

\begin{figure}
	\centering
	\includegraphics[width=\textwidth]{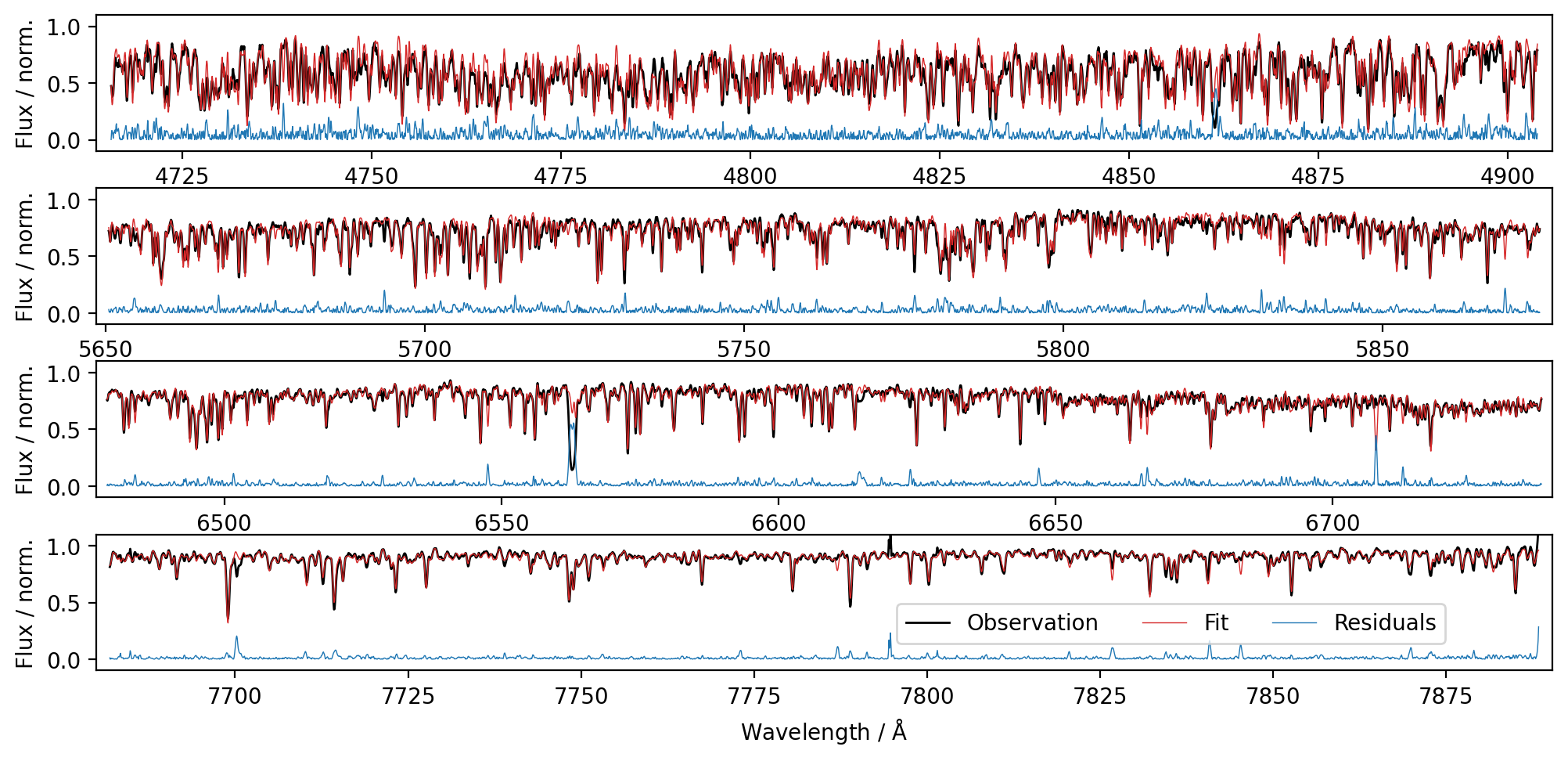}
	\caption{The observed (black) and fitted (red) synthetic HERMES spectra from GALAH DR4, showing generally good agreement and small residuals (blue) across the entire spectrum. While some individual lines are deviating, the overall fit is well aligned with the observations, whose continuum is suppressed by molecular absorption features across the entire wavelength region of this luminous cool giant star.}
	\label{fig:hermes_spectrum}
\end{figure}

The peak of the light curve (Figure~\ref{fig:lc}) is flattened compared to the PSPL model, suggesting possible finite-source effects \citep{Gould1994a}. We fit the FSPL model to the light curve by introducing an additional parameter, $\rho = {\theta_*}/{\theta_\E}$. In the FSPL model, $A_{\rm FSPL}$ is calculated by integrating $A_{\rm PSPL}$ over a limb-darkened source disk. Using the best-fit spectroscopic parameters (see \S~\ref{sec:spectrum}), we estimate the limb-darkening coefficients {$(\Gamma_{g^\prime}, \Gamma_{r^\prime}, \Gamma_{i^\prime}, \Gamma_G, \Gamma_K)= (0.91, 0.77, 0.63, 0.74, 0.29)$} based on \citet{Claret2011} and \citet{Claret2019}.

\begin{figure}[htbp]
	\centering
	\includegraphics[width=1\textwidth]{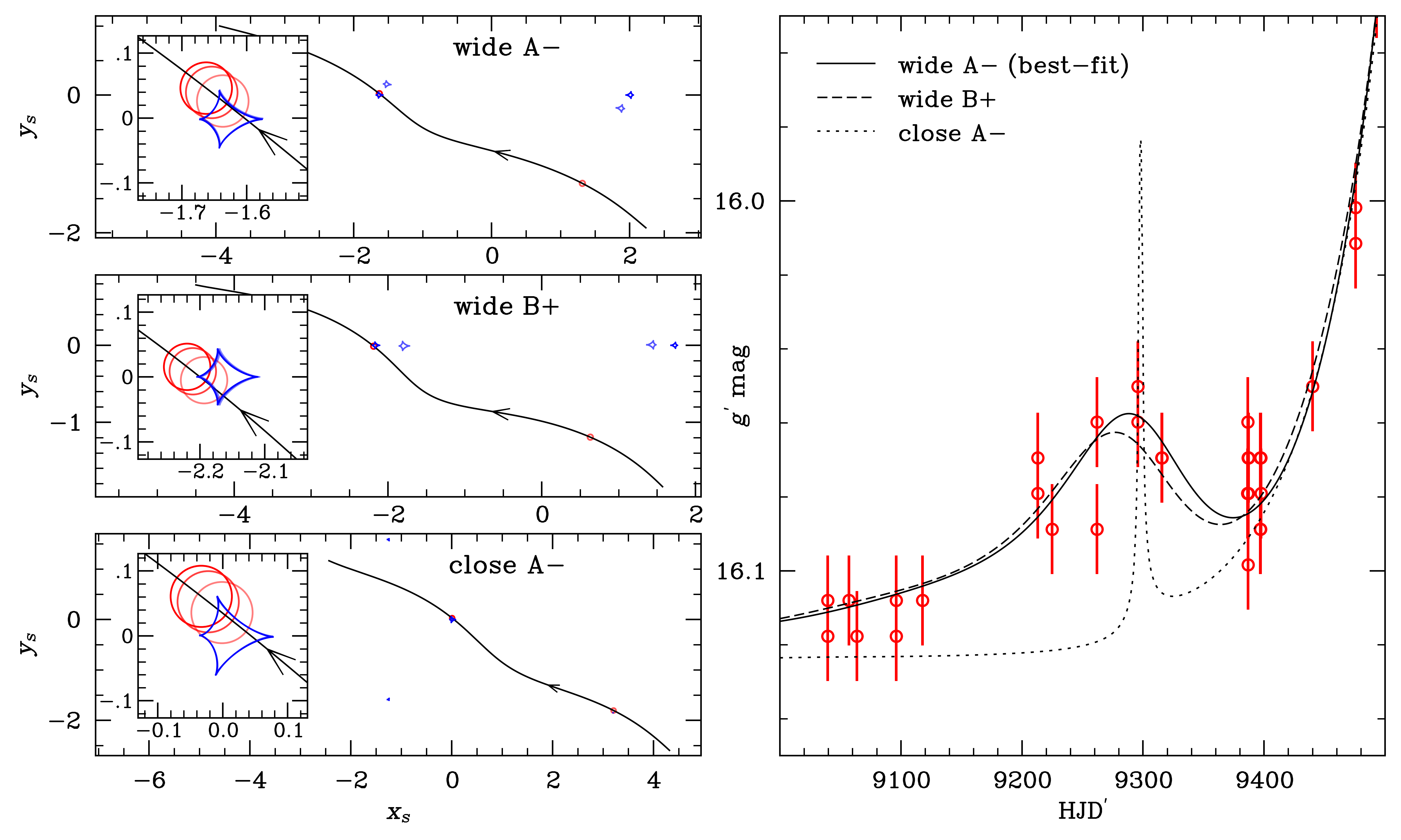}
	\caption{The source trajectories and light curves of three 2L1S models (\wide{}, ``wide $B+$'', and \close{}) taking the microlens parallax and binary orbital motion into account. Left:
		the upper, middle, and lower subpanels show the caustics (blue) and source trajectories (black curves with arrows) for the ``wide A$-$'' (best-fit), ``wide B$+$'', and ``close A$-$'' solutions, respectively. In each subpanel, we show the source positions (red circles) and caustics at $\rm HJD'=9300$ and $9601$, with increased opacity for the latter. The inset of each subpanel displays the source positions and caustics on the three nights of the VLTI GRAVITY observations. Right: the {Gaia} $G$-band light curve shows a broad feature at $\rm HJD' \sim 9300$, and the three best-fit 2L1S models are shown with solid (wide A$-$), dashed (wide B$+$), and {dotted} (close A$-$) lines, respectively. The wide solutions can match this feature well, while the ``close A$-$'' solution cannot. {The sharply peaked spike of the ``close A$-$'' solution is induced by the source crossing a planetary caustic due to dramatic orbital motions, but even this finely tuned configuration fails to reproduce the observed feature.}}
	\label{fig:traj}
\end{figure}

The best-fit FSPL model is shown as a dashed black line in the top panel of Figure~\ref{fig:lc}, and the residuals are shown in the bottom panel. The residuals exhibit significant trends deviating from the FSPL model, spanning $\sim 10$ days around the peak (at {$\rm HJD' \sim 9600$}). These systematic deviations cannot be absorbed by adjusting either the limb-darkening coefficients or the free parameters in the FSPL models.  {Deviations from single-lens models suggest the possible existence of a companion to the lens or the source. There is also a $\sim 0.1$\,mag broad bump-like feature in the {Gaia} $G$-band light curve at around $\sim300$\,days before the peak  (see the right panel of Figure~\ref{fig:traj}). Such an additional feature excludes single-lens binary-source models, and we study binary-lens models in the following.}

\subsection{Binary-lens Model}
\label{sec:binarymodel}
The light curve can be well described by the binary-lens single-source (2L1S) model.
We start the analysis with the static binary-lens model, which includes three additional parameters $(s, q, \alpha)$ to describe a nonrotating binary-lens system: $s$ is the projected angular separation of the binary components normalized by $\theta_\E$, $q$ is the binary mass ratio, and $\alpha$ is the angle between the source-lens trajectory and the binary-lens axis.
We use the advanced contour integration package \texttt{VBBinaryLensing} \citep{VBBL2010, VBBL2018} to calculate the binary-lens magnification $A(t, \Theta_{\rm 2L 1S})$, given the parameters $\Theta_{\rm 2L 1S} = \{s, q, \alpha, \rho , t_0, u_0, t_\E\}$.

\begin{figure}[htbp]
	\centering
	\includegraphics[width=0.8\textwidth]{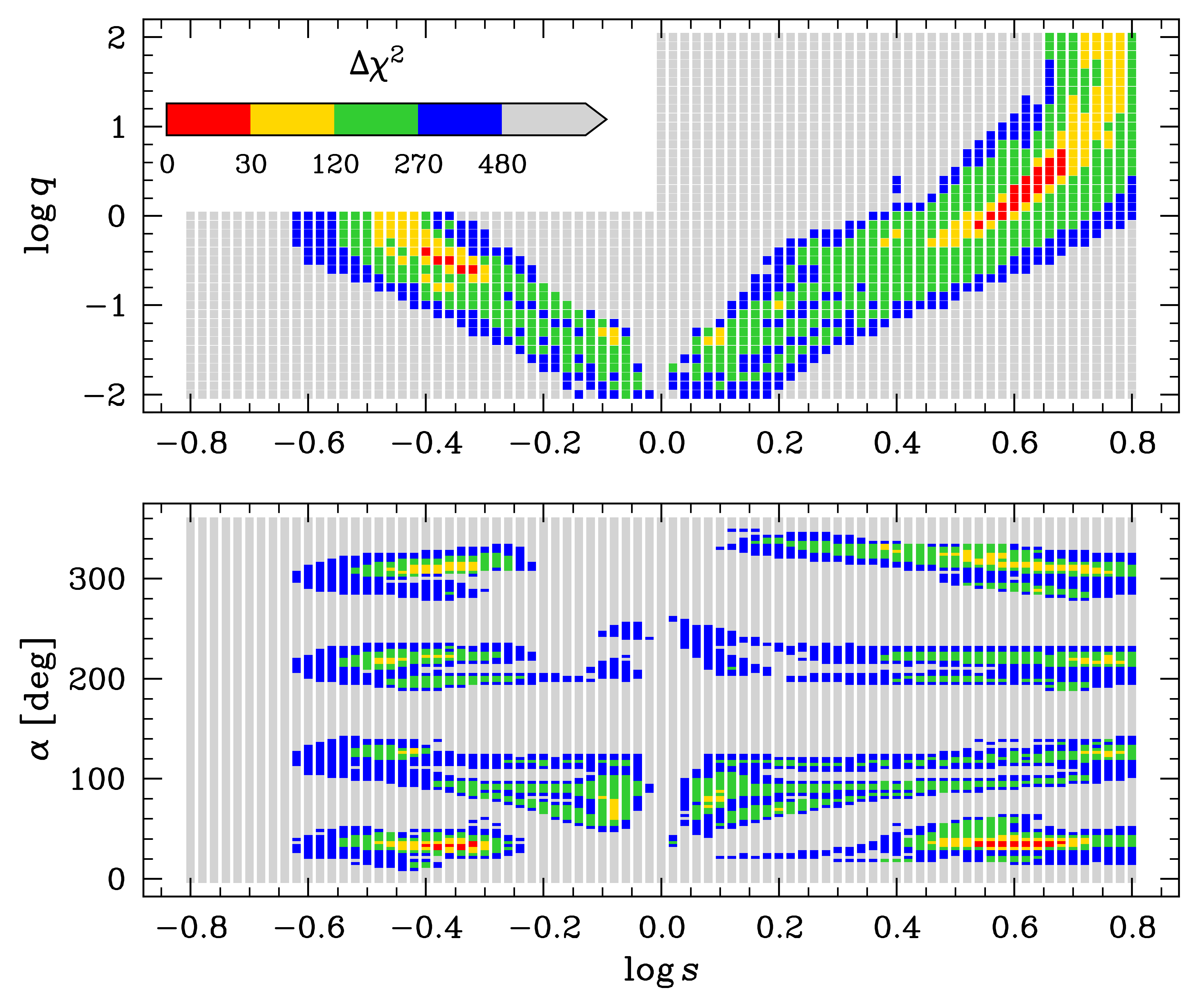}
	\caption{$\Delta \chi^2$ map for the grid search of 2L1S light-curve models, in the $\log s-\log q$ (upper panel) and $\log s-\alpha$ (lower panel) planes. For $\log s < 0$, there is a perfect degeneracy $(\log q ,\alpha) \to (-\log q, \alpha + \pi)$, and thus the search is only conducted with $\log q\le 0$. The regions with $\Delta \chi^2 < 30, 120, 270, 480$, and $>480$ with respect to the best fit are color-coded with red, yellow, green, blue, and light gray, respectively.}
	\label{fig:grid_search}
\end{figure}

We start by scanning the seven-dimensional parameter space on a fixed $(\log s, \log q, \alpha)$ grid with $-1 \le \log s \le 1$, $-2 \le \log q \le 1$  and $0  \le\alpha  < 2\pi$. At each grid point, we minimize the $\chi^2$ by setting the four remaining parameters $(t_0, u_0, t_\eff\equiv u_0 t_\E, t_* \equiv \rho t_\E)$ free in Markov Chain Monte Carlo (MCMC) using the \texttt{EMCEE} package \citep{Foreman-Mackey2013}. The resulting $\chi^2$ map is shown in Figure~\ref{fig:grid_search}, from which we identify five local minima in the close-binary $(s<1)$ regime and nine local minima in the wide-binary $(s>1)$ regime. The severe finite-source effects smear out the sharp magnification structure of caustics, leading to a plethora of local minima over a broad range of $\log q$.
For close solutions, there is a perfect degeneracy of $(\log q, \alpha) \to (-\log q, \alpha + \pi)$, and therefore we only conduct the grid search with  $\log q\le 0$.
For wide solutions, we have a fourfold degeneracy in $\alpha$. When $q \gg 1$, the ``planetary'' caustic associated with secondary mass is reduced to a Chang-Refsdal \citep{ChangRefsdal1979} caustic with shear $s^{-2}/q \ll 1$. The four identical cusps of Chang-Refsdal lead to a degeneracy of $\alpha \to \alpha + \pi/2$, as shown in Figure~\ref{fig:grid_search}. The solutions with $q \gg 1$ can correspond to an unrealistic value of $t_\E$. We limit our analysis to $\log q < 3$, which corresponds to an upper limit of $t_\E \sim 2000\,$days.

Then we probe all local minima by setting all the parameters free in MCMC. We also introduce higher-order effects into the modeling. We include the annual microlens parallax effects \citep{Gould1992}, which are the light-curve distortions due to the Earth's orbital acceleration toward the Sun. The static binary-lens model has a perfect geometric degeneracy $(u_0, \alpha) \to -(u_0, \alpha)$, which is broken when taking the microlens parallax into account.
For the $u_0>0$ and $u_0<0$ geometry, the lens takes different sides of the source trajectory, leading to different magnifications. {As defined in \citet{Gould2004}, the microlens parallax vector $\bpi _\E$ is described by the east and north components $\bpi _\E = (\pi _{\rm E,E}, \pi_{\rm E,N})$.}
We also consider the orbital motion of the binary lens, which can have degeneracies with microlens parallax  \citep{Batista2011, Skowron2011}.
We apply the linearized orbital motion
\begin{equation}
	s(t) = s_0+\frac{ds}{dt}(t-t_0);\;\;\; \alpha(t)  = \alpha_0 + \frac{d\alpha }{dt}(t-t_0).
\end{equation}
with the two parameters $ds/dt$ and $d\alpha/dt$.
This approximation is subject to physical constraints, as discussed in \citet{Dong2009}. For a bound system, the ratio of the projected kinetic to the potential energy $\beta$ should be less than unity:
\begin{equation}
	\beta  = \left\vert \frac{\rm KE_\perp}{\rm PE_\perp}\right\vert = \frac{(\gamma \cdot\yr) ^2 }{8\pi ^2 }\frac{M_\sun\kappa \pi _\E}{\theta _\E}\left(\frac{s }{\pi _\E + \pi _\Sc/\theta _\E}\right)^3 < 1,
\end{equation}
where $\gamma = (ds/s\,dt, d\alpha /dt)$ and $\pi _\Sc$ is the trigonometric parallax of the source star.
We adopt $\theta_\E = 0.73\,\mas$ and $\pi _\Sc = 0.185\,\mas$ (see \S~\ref{subsec:source_param}) and restrict the MCMC exploration to $\beta < 1$.

We find that the best-fit solution is a wide binary with $u_0<0$ and we label it as \wide{}. The source trajectory and caustics are shown in the upper left subpanel of Figure~\ref{fig:traj}. The \wide{} solution has $\chi^2 = 948.9$, and all other solutions are worse by $\Delta \chi^2 > 50$. Nevertheless, we keep all solutions with $\Delta \chi^2 < 81$ (listed in Table~\ref{tab:lc_param}) for further analysis incorporating the VLTI GRAVITY data.

\section{Interferometric Data Analysis}\label{subsec:vlti_analysis}
\subsection{Parameterization}
The images generated from the binary-lens equation are oriented with respect to the binary-lens axis and in the units of the Einstein radius, so they need to be rotated to the sky plane and scaled by $\theta_\E$ to compute the closure phases. We define $\Psi$ as the position angle of the binary-lens axis. The direction of the lens-source relative motion $\bmu_\rel$ , which is the same as that of the microlens parallax $\bpi _\E$, can be expressed as
\begin{equation}
	\label{eq:piE_direction}
	\Phi _{\bpi} = \alpha + \Psi
\end{equation}
We incorporate the two parameters $\{\theta_\E, \Psi\}$ in addition to the 2L1S parameters $\{s, q, \alpha, \rho, t_0, u_0, t_\E\}$ for modeling the interferometric data. The microlens parallax is reparameterized by $(\pi_{\rm E, N}, \pi _{\rm E, E}) = \pi _\E \cdot (\cos \Phi _{\bpi}, \sin \Phi _{\bpi})$ with $\Phi _{\bpi} = \alpha + \Psi$.

\subsection{Probing Interferometric Observables}
\label{subsec:vlti_observables}
Prior to carrying out the full joint analysis of VLTI and light curve, we first investigate the VLTI parameters with limited input from the light-curve models.
We start the investigation using only $(s, q, \rho)$ from the best-fit light-curve model ``wide A$-$.'' We perform a four-dimensional grid search on $(\theta_\E, \Psi, u_x, u_y)$ to fit the closure-phase data from each of the eight VLTI exposures, where $(u_x, u_y)$ are the source's coordinates on the binary-lens plane.
At each grid point, we generate the images scaled and rotated to the sky plane based on $(\theta_\E, \Psi)$ and then calculate the closure phases and corresponding $\chi_{{\rm VLTI}, i}^2(\theta_\E, \Psi, u_x, u_y)$ for every VLTI exposure $i$. We describe our method of calculating the closure phase from the binary-lens images in Appendix~\ref{appendix:binaryimage}. Our closure-phase calculations from the images are validated with the PMOIRED software\footnote{\url{https://github.com/amerand/PMOIRED}} \citep{pmoired}.

To exclude unreasonable source positions, we add a penalty term,
\begin{equation}
	\chi^2_{i} = \chi_{{\rm VLTI}, i}^2 + \left(\frac{A(u_{x,i}, u_{y,i}) - A_{{\rm LC},i} }{0.1 A_{{\rm LC},i}}\right)^2,
	\label{eq:penality}
\end{equation}
for the deviation from light-curve magnification for each VLTI exposure $i$, where $A_{{\rm LC},i}$ is the magnification from the \wide{} model at the $i$th VLTI exposure, and we adopt a fractional error of  $10\%$ so that $(u_{x,i}, u_{y,i})$ are only subject to loose light-curve magnification constraints.

A self-consistent VLTI model should have the same values of $(\theta_\E, \Psi)$ on all the exposures. Therefore, for each set of $(\theta_\E, \Psi)$, we calculate
\begin{equation}
	\chi_{\rm tot}^2 (\theta_\E, \Psi) = \sum_i \min \left[\chi_{i}^2(\theta_\E, \Psi, u_{x, i}, u_{y, i})\right],
\end{equation}
where $\min \left[\chi_{i}^2(\theta_\E, \Psi, u_{x, i}, u_{y, i})\right]$ is the minimum $\chi^2$ for a given $(\theta_\E, \Psi)$ in the $i$-th VLTI exposure.

\begin{sidewaysfigure}
	\subfloat{
		\includegraphics[width=0.5\linewidth]{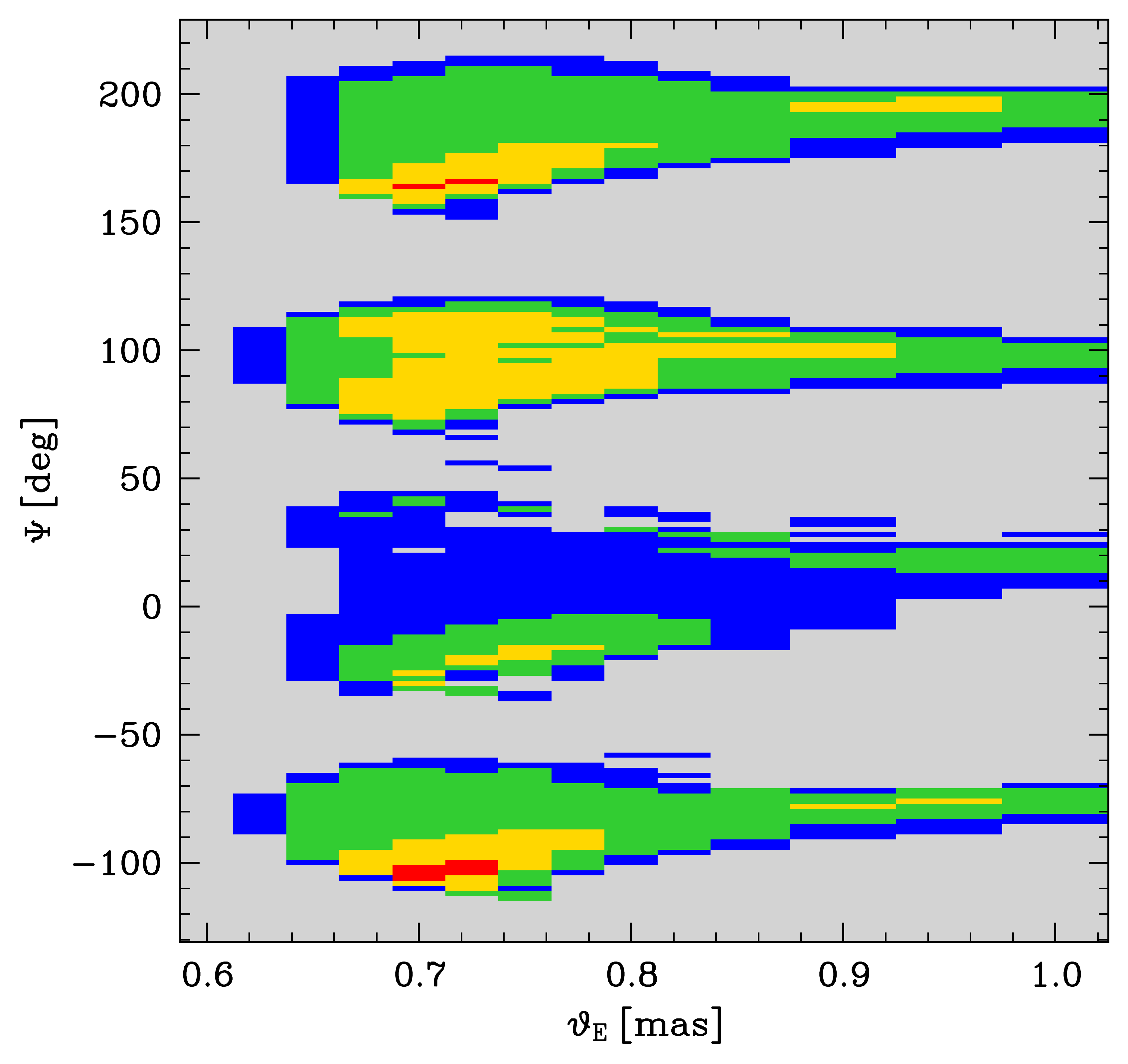}
	}
	\subfloat{
		\includegraphics[width=0.5\linewidth]{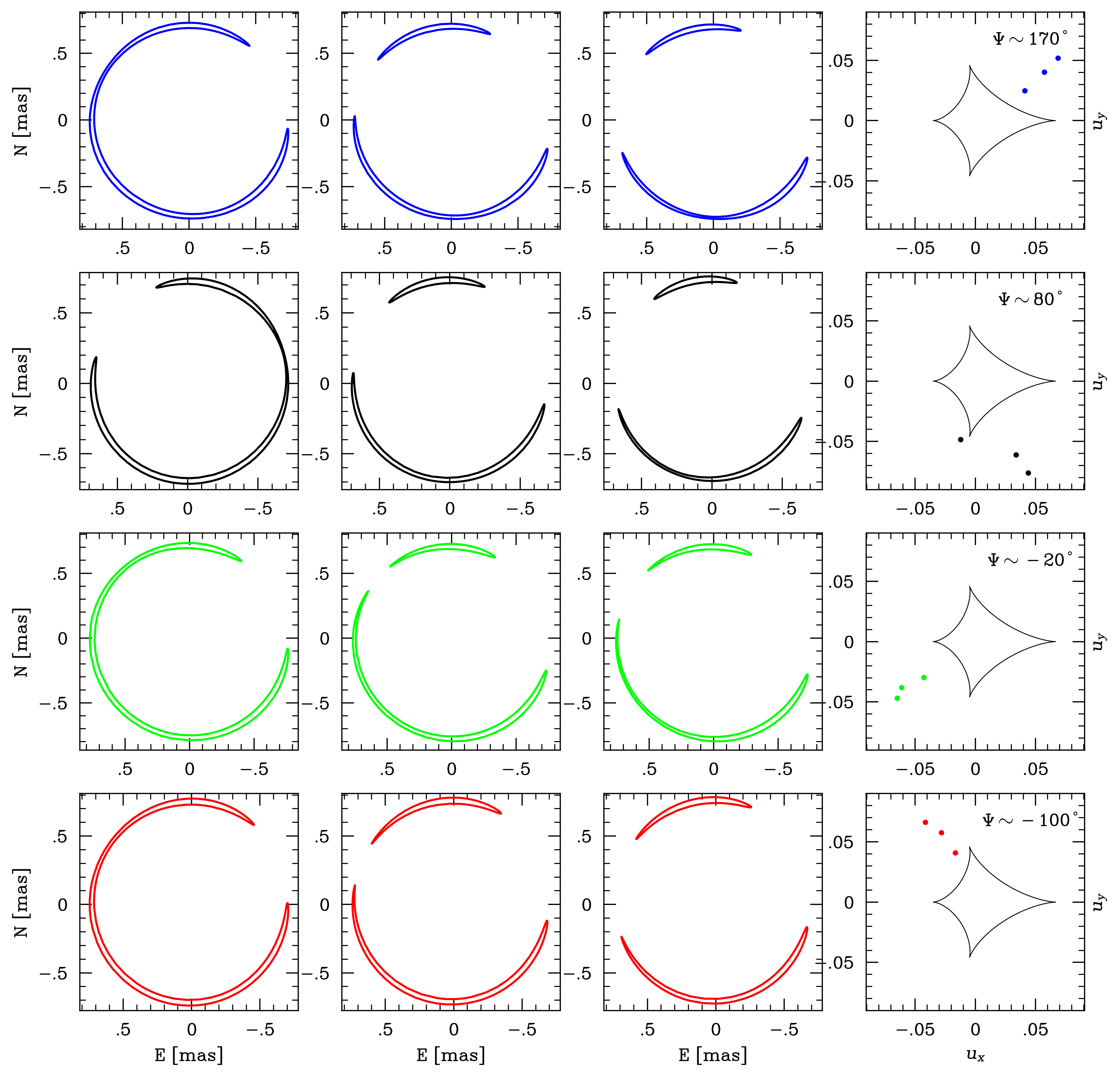}
	}
	\caption{Left: the $\Delta \chi_{\rm tot}^2(\theta_\E, \Psi)$ map from fitting the VLTI data evaluated at $(s, q, \rho) = (3.8, 0.84, 0.044)$. Areas with $\Delta \chi ^2 < 30, 120, 270, 480$, and $\Delta \chi ^2  > 480$ are color-coded in red, yellow, green, blue, and gray, respectively. There are four local minima around $\Psi = (170\deg, 80\deg, -10\deg,  -100\deg)$. Right: images and geometry of the four local minima. The four rows from top to bottom correspond to local minima at $\Psi = 170\deg\,(\rm blue), 80\deg \,(\rm black), -10\deg \,(\rm green),$ and $-100\deg\,(\rm red)$, respectively. The first three columns show the model images oriented in the east and north directions on the three VLTI nights. The last column shows the source positions (solid dots) on the three VLTI nights and the caustics (black closed curve).}
	\label{fig:thetaE_psi}
\end{sidewaysfigure}

The left panel of Figure~\ref{fig:thetaE_psi} shows the $\Delta \chi^2_{\rm tot}(\theta_\E, \Psi)$ map, from which we identify four local minima. The $\theta_\E$ is within a broad range of $0.6 - 1.0\,\mas$, and $\Psi$ is subject to a fourfold degeneracy around $(-100\deg, -10\deg, 80\deg, 170\deg)$, each with an uncertainty of $\sim 10\deg$.
The fourfold degeneracy corresponds to having the source positions near the four cusps of the ``central'' caustic associated with the primary mass, resulting in similar image morphologies that are compatible with the VLTI data (see the subpanels on the right of Figure~\ref{fig:thetaE_psi}).

\begin{figure}[htbp]
	\centering
	\includegraphics[width=0.8\textwidth]{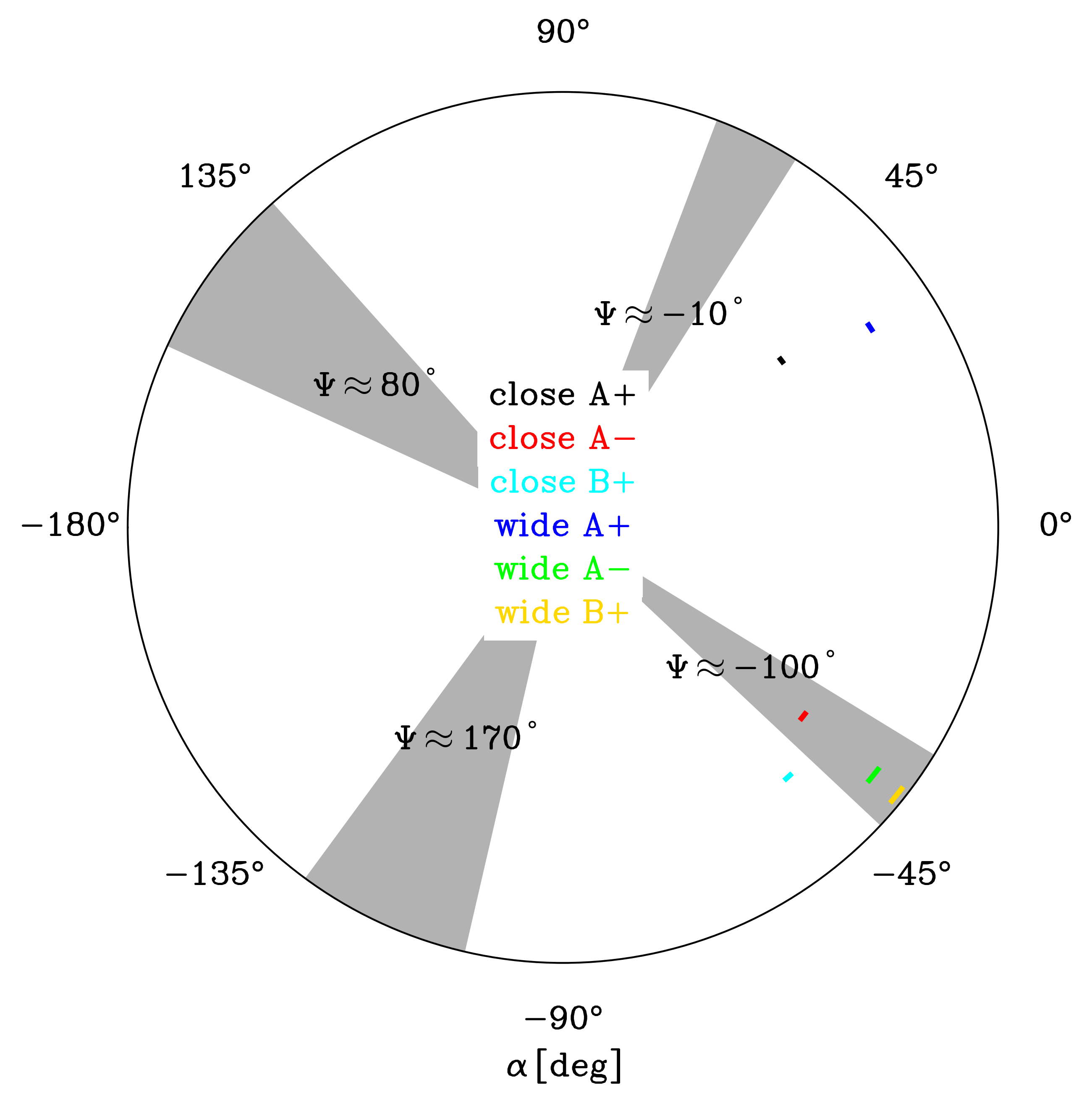}
	\caption{The comparisons of $\alpha_{\rm LC}$ derived from light-curve fitting with $\alpha_{\rm VLTI}$ inferred from VLTI data. The colored arcs denote the 1$\sigma$ range of $\alpha_{\rm LC}$ for the solutions ``close A$+$'' (black), ``close A$-$'' (red), ``close B$+$'' (cyan), ``wide A$+$'' (blue), ``wide A$-$'' (green), and ``wide B$+$'' (yellow), respectively. The shaded regions indicate the 3$\sigma$ range of $\alpha_{\rm VLTI}$ at $\Psi \sim -100\deg, -10\deg, 80\deg$, and $170\deg$. Three solutions (``wide A$-$'', ``wide B$+$'' and ``close A$-$'') have consistent $\alpha_{\rm LC}$ with $\alpha_{\rm VLTI}$.}
	\label{fig:alpha}
\end{figure}

Next, we further probe the four local minima by jointly fitting all epochs of VLTI data. In a satisfactory VLTI microlensing model, the source positions should lie approximately on a straight line, following the source-lens relative motion, so we parameterize the source positions $(u_{x}, u_{y})$ with $(u_0, t_0, t_\E, \alpha)$ as
\begin{eqnarray*}
	u_{x} &= u_0 \sin \alpha - \tau \cos \alpha,\\
	u_{y} &= -u_0 \cos \alpha - \tau \sin \alpha,
\end{eqnarray*}
where $\tau  = (t -t_0) / {t_\E}$.  The interferometric parameters $(\theta_\E, \Psi)$ are seeded from the abovementioned four local minima. We allow $(\log s, \log q, \log \rho)$ to freely vary, and we no longer add the $\chi^2$ penalty term as in Eq.~\ref{eq:penality}.  The four best-fit solutions have similar $\chi^2$ ($\Delta\chi^2<10$), so the VLTI data alone cannot distinguish between them. All the solutions strongly prefer $u_0<0$. As in the previous single-lens events, $u_0>0$ and $u_0<0$ solutions can be distinguished with two or more epochs of VLTI data  \citep{Dong2019, Zang2020, Cassan2022}. The four degenerate solutions correspond to four distinct $\alpha$ values, denoted as $\alpha _{\rm VLTI}$, which are displayed as the shaded regions in Figure~\ref{fig:alpha}. The allowed $(\log s, \log q)$ from the posteriors have broad ranges, suggesting that the resulting constraints on $\alpha _{\rm VLTI}$ are not specific to the seed solution of ``wide A$-$,'' so we use them as the starting point for the joint analysis with the light curves in  \S~\ref{subsec:joint_analysis}.

\subsection{Joint Analysis with Light Curves}
\label{subsec:joint_analysis}
\begin{figure}[htbp]
	\centering
	\includegraphics[width=0.8\textwidth]{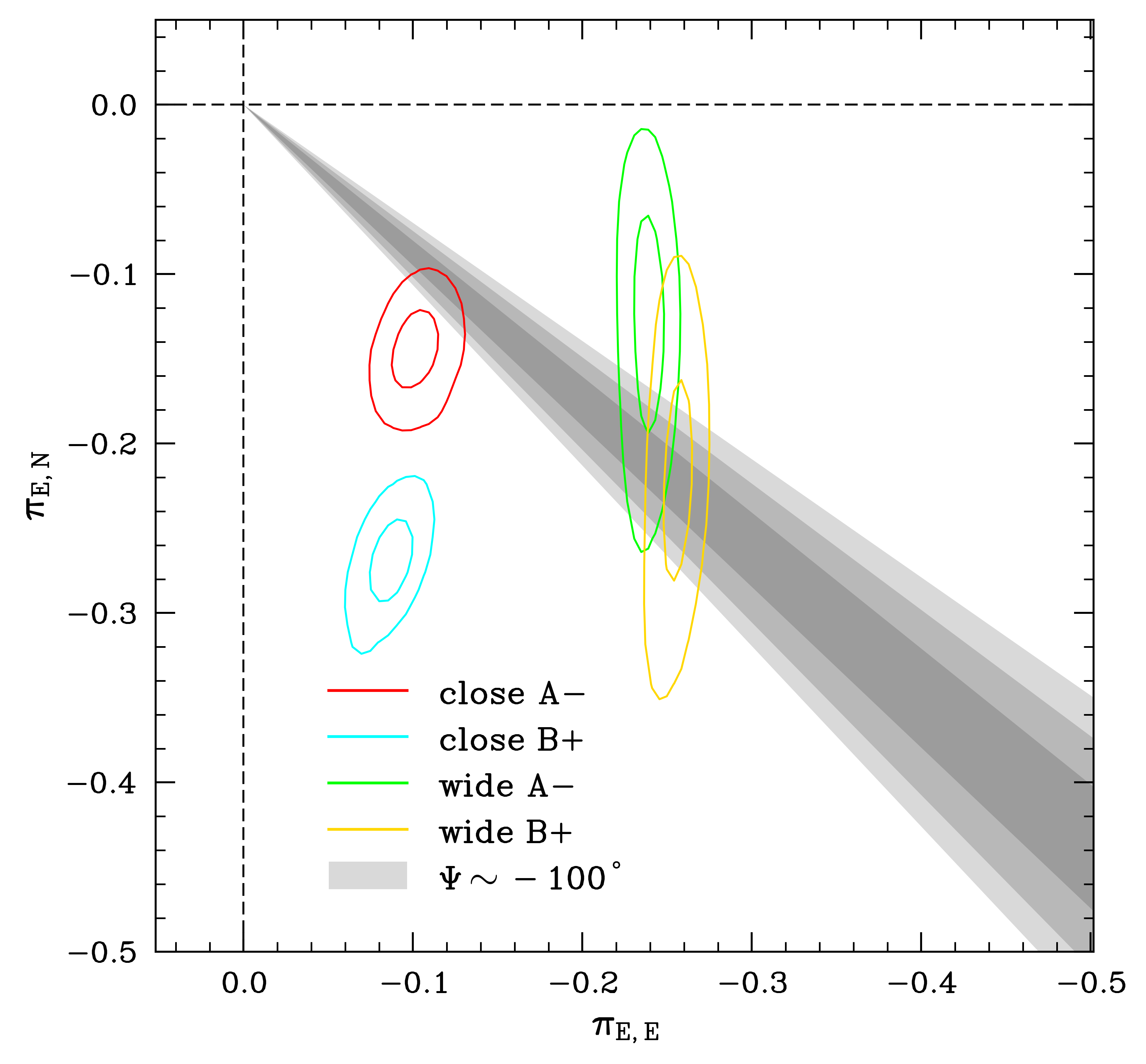}
	\caption{The 1 and 2$\sigma$ posterior distributions of the microlens parallax vector $\bpi_\E = (\pi_{\E,\E}, \pi_{\E,{\rm N}})$ from light-curve analysis for the ``close A$+$'' (red contours), ``close B$+$'' (cyan contours), ``wide A$-$'' (green contours), and ``wide B$+$'' (yellow contours) solutions, respectively. The shaded regions are the 1$\sigma$, 2$\sigma$, and 3$\sigma$ constraints on $\Phi_{\pi}$ derived from the interferometric data for the local minimum of $\Psi\sim-100\deg$. The light-curve and VLTI constraints are consistent for the ``wide A$-$'', ``wide B$+$'' and ``close A$-$'' solutions. }
	\label{fig:piE}
\end{figure}

In this section, we present the joint analysis of VLTI and light-curve modeling.
We begin by comparing the light-curve parameters and VLTI constraints. Figure~\ref{fig:alpha} shows the comparisons between the $\alpha$ parameters $\alpha_{\rm LC}$ derived from the light-curve solutions with $\alpha _{\rm VLTI}$. The three solutions are mutually compatible within $3\sigma$, and they are all at $\Psi \sim -100 \deg$. These include one close solution (\close{}) and two wide solutions (\wide{} and ``wide B$+$''). Next we check the consistency of the direction $\bpi_\E$ between the light-curve solutions and VLTI solution at $\Psi \sim -100\deg$. Figure~\ref{fig:piE} displays the posterior distribution of $(\pi_{\rm E, E}, \pi_{\rm E, N})$ and the $\Phi _{\bpi}$ constraint from VLTI. For three light-curve solutions (\wide{}, ``wide B$+$'', and \close{}), the two directions agree to $\sim 1\sigma$.

Then we perform a joint fit to the VLTI and light-curve data for the \wide{}, ``wide B$+$'', and \close{} solutions. We include the higher-order effects of the microlens parallax and binary orbital motion.
	{In addition, we account for the limb-darkening effects in the closure-phase modeling by adopting $\Gamma_{K}=0.29$, based on our adopted source stellar parameters (\S~\ref{subsec:espl_analysis}). Compared to the uniform-brightness profile, limb darkening presents minor improvements in the goodness of fit, with $\Delta \chi^2 \sim (3, 7, 8)$ for the solutions (wide A$-$, wide B$+$, close A$-$), respectively, and the best-fit parameters with and without the limb-darkening effects are consistent with each other. Thus, the non-limb-darkening approximation adopted in the closure-phase analysis is adequate for probing local minima with grid searches (\S~\ref{subsec:vlti_observables}). Nevertheless, we consider the limb-darkening effects to derive the final results from the joint fit.}

\begin{figure}[htbp]
	\subfloat{
		\includegraphics[width=0.45\linewidth]{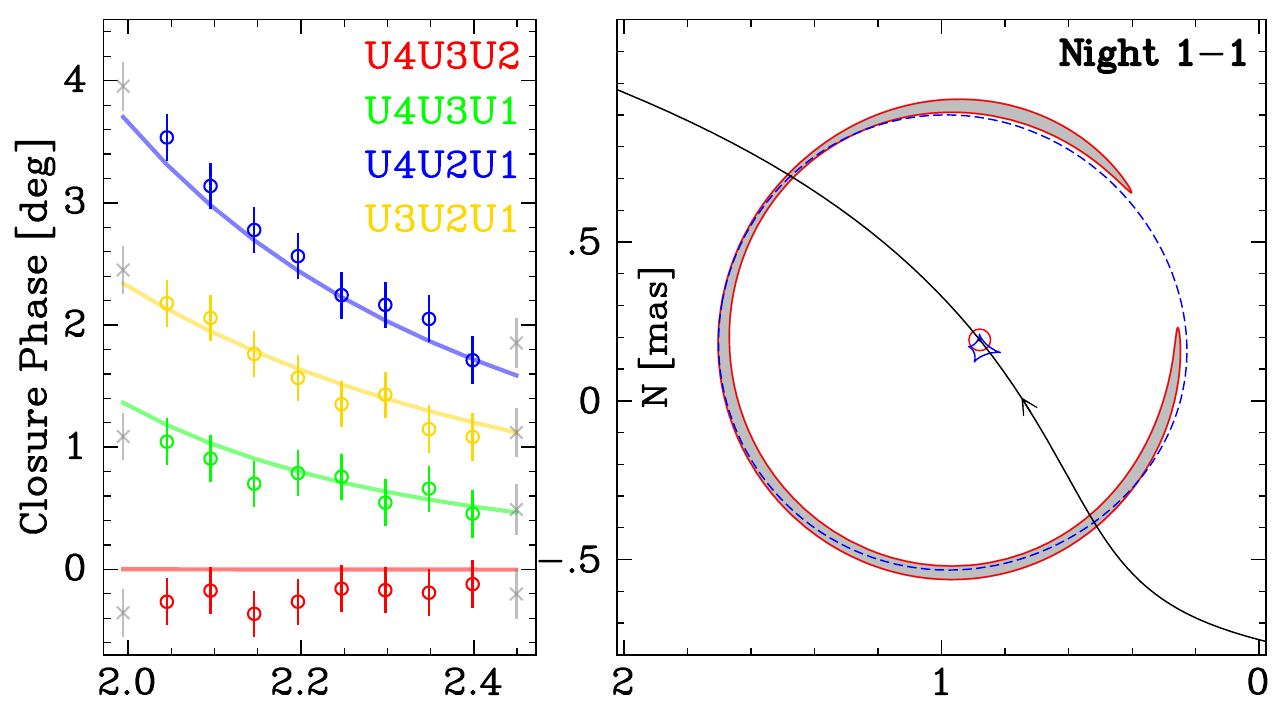}
	}
	\subfloat{
		\includegraphics[width=0.45\linewidth]{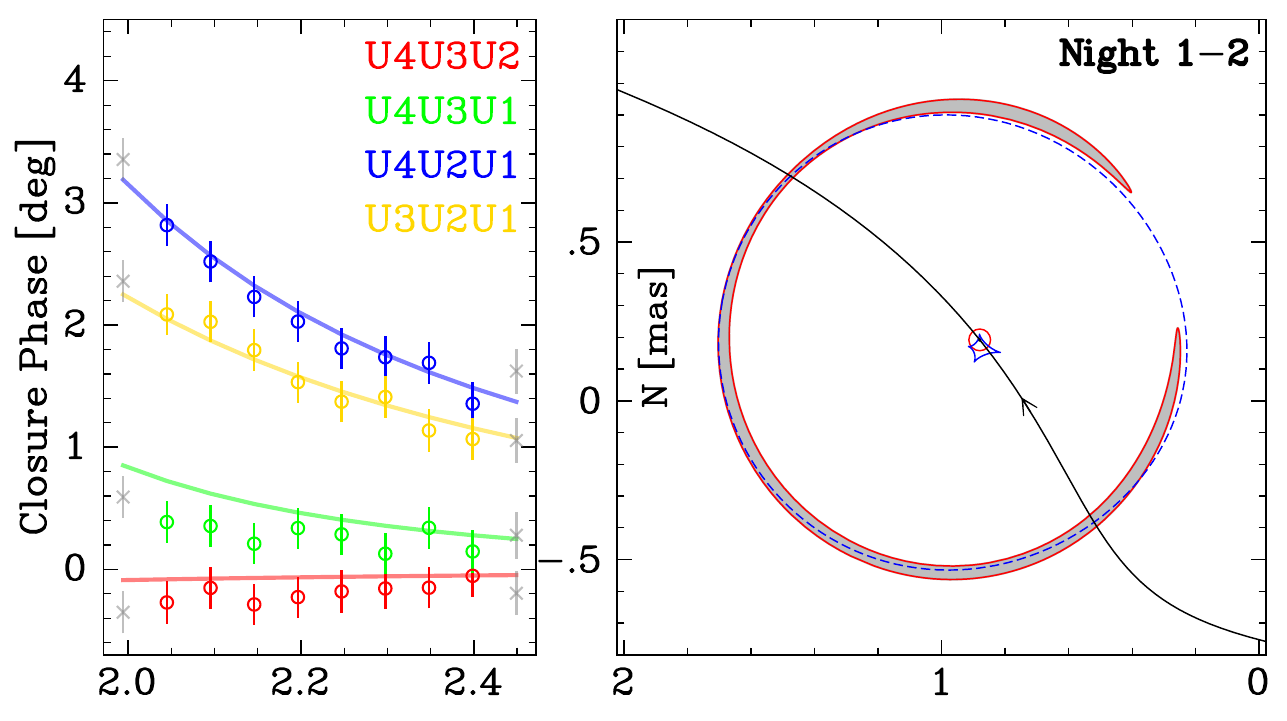}
	}\\
	\subfloat{
		\includegraphics[width=0.45\linewidth]{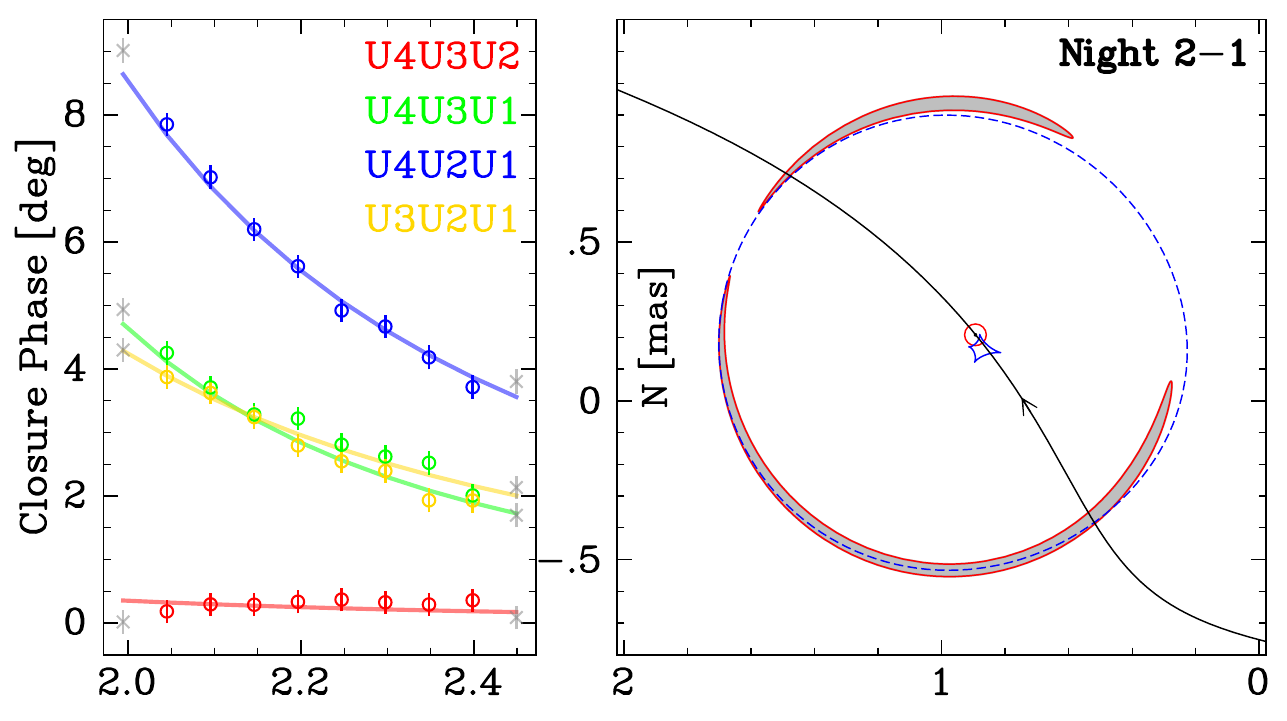}
	}
	\subfloat{
		\includegraphics[width=0.45\linewidth]{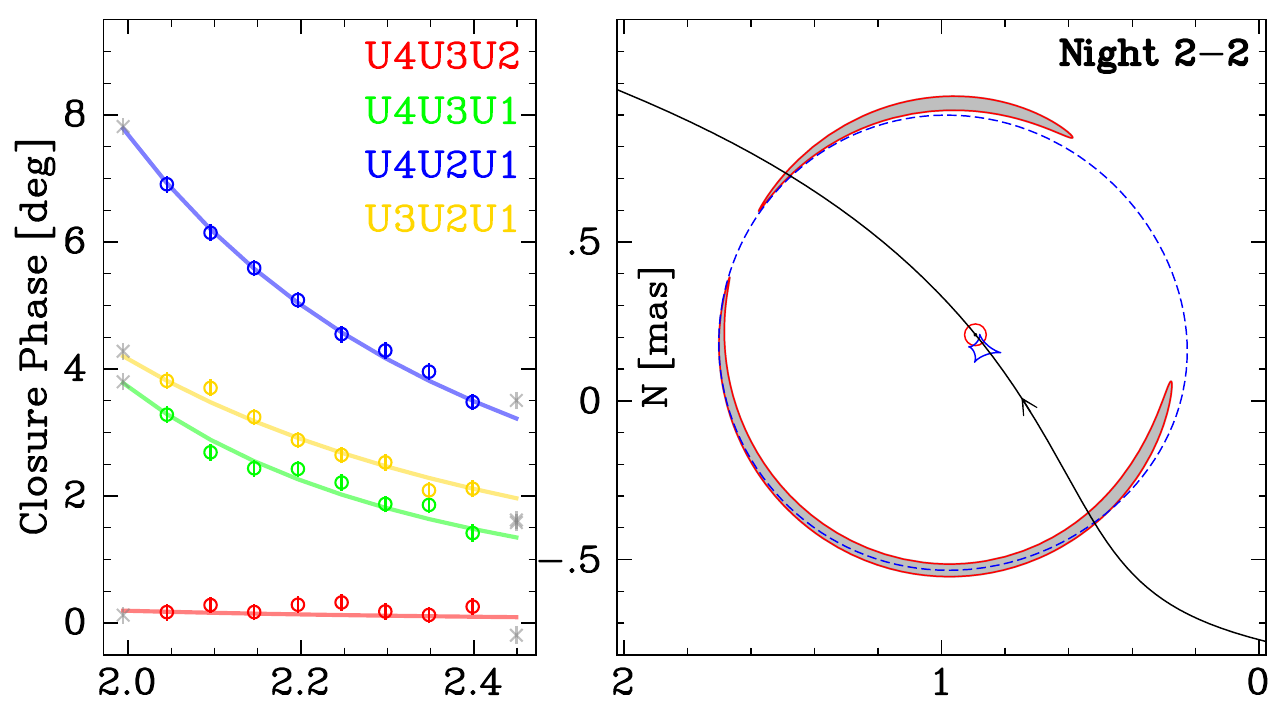}
	}\\
	\subfloat{
		\includegraphics[width=0.45\linewidth]{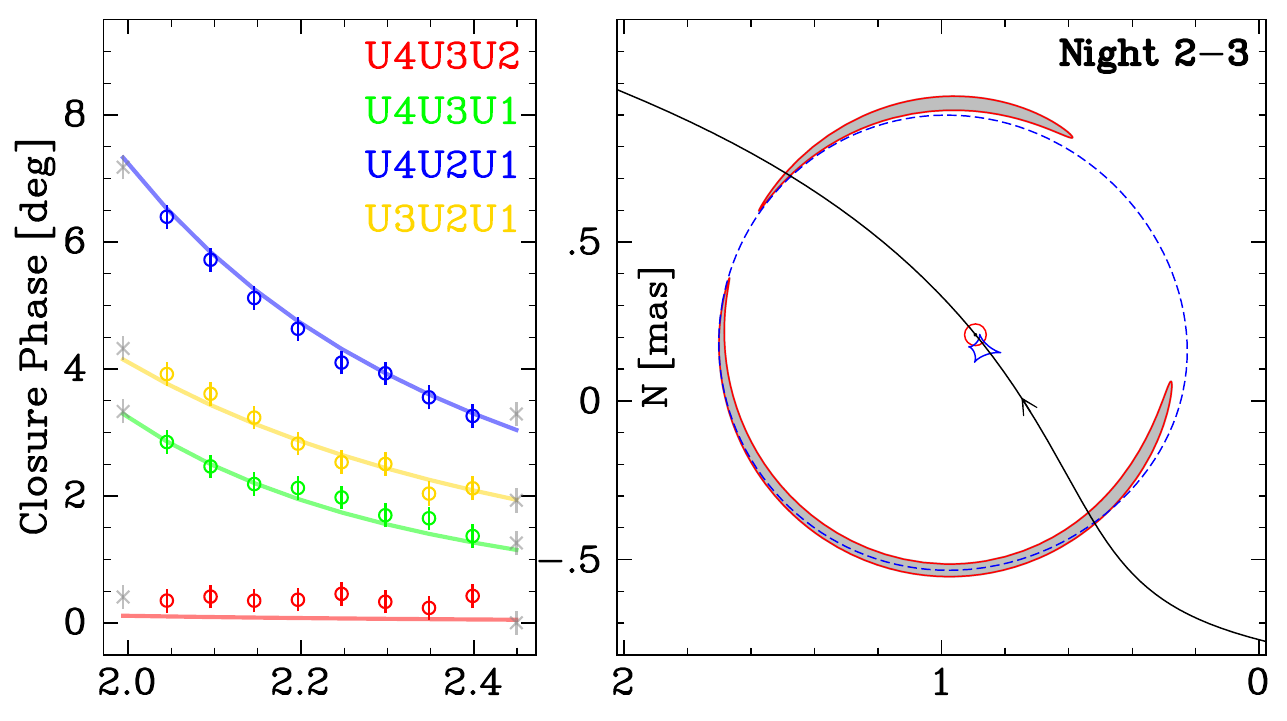}
	}
	\subfloat{
		\includegraphics[width=0.45\linewidth]{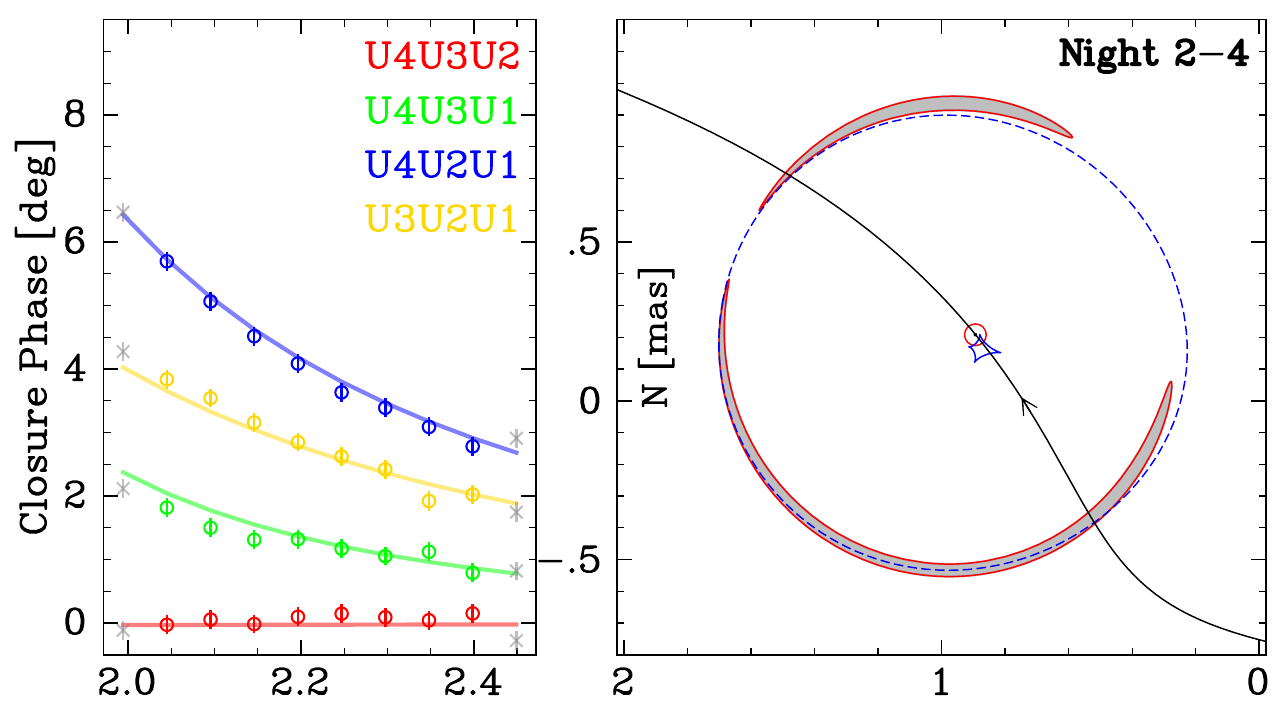}
	}\\
	\subfloat{
		\includegraphics[width=0.45\linewidth]{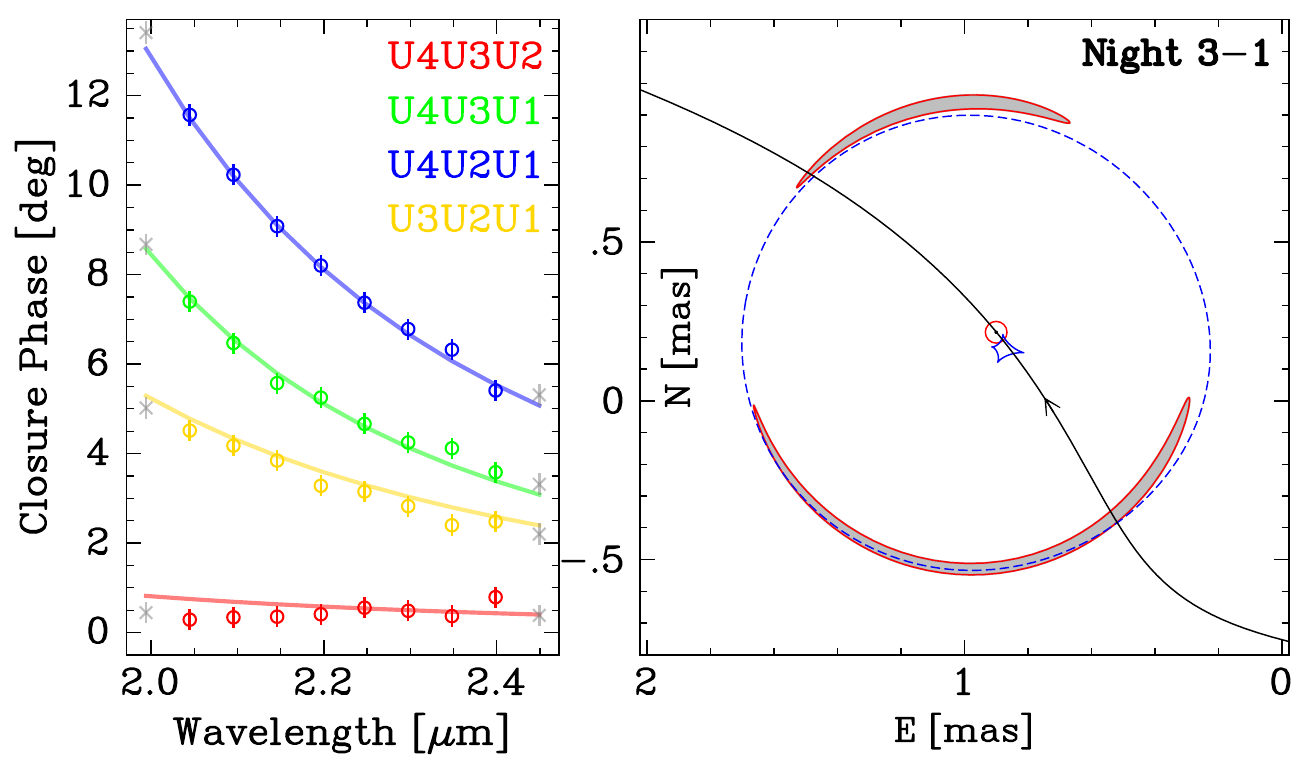}
	}
	\subfloat{
		\includegraphics[width=0.45\linewidth]{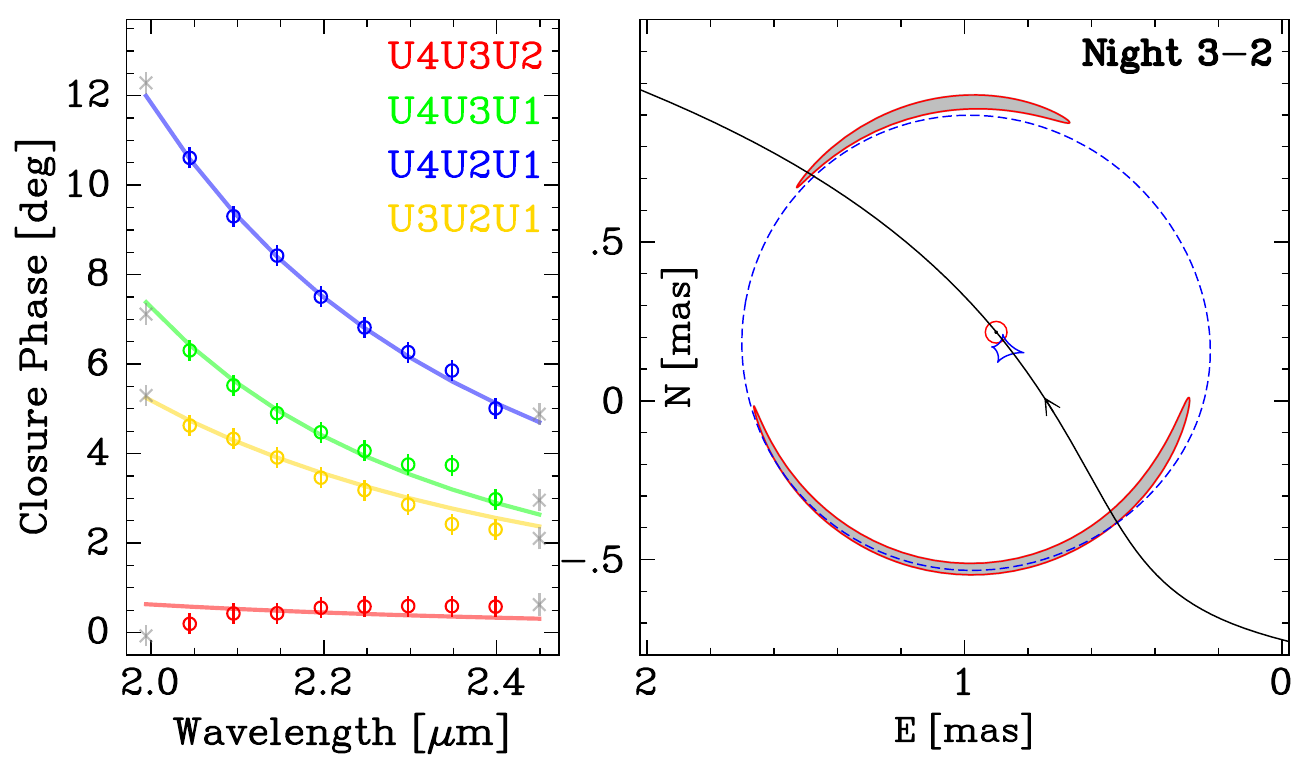}
	}
	\caption{Four $\times$ two panel groups showing the closure-phase data and the best-fit microlensing model (\wide{}) for the eight VLTI GRAVITY exposures obtained over three nights. Within each group, the left subpanel shows the closure-phase data as a function of wavelength, with the best-fit model overplotted as solid lines; the right subpanel shows the caustics (blue), source trajectory (black line with an arrow), and source position (red circle) at the time of the VLTI observation, with the microlensed images displayed as shaded regions and the critical curves as blue dashed lines.}
	\label{fig:cpfit}
\end{figure}

    Table~\ref{tab:joint_param} lists the best-fit parameters and their uncertainties. The global best fit is the \wide{} solution, and Figure~\ref{fig:cpfit} shows its best-fit closure-phase models along with the geometric configurations of the source, caustic, and images. The \close{} solution is worse by {$\Delta \chi^2 \approx45$}, which is somewhat ({$\approx25$}) less compared with the light-curve-only $\Delta \chi^2$. Therefore, the preference for the \wide{} over the \close{} comes from the light-curve fitting. {It mainly stems from the $\sim 0.1$\,mag broad bump around $\rm HJD' \sim 9300$ in the {Gaia} light curve, as mentioned in \S~\ref{subsec:espl_analysis}. This feature can be well matched by the wide solutions. In contrast, for the ``close A$-$'' solutions, the orbital motion parameters can be tuned to induce source crossing of a planetary caustic at $\rm HJD' \sim 9300$, but the corresponding perturbation is too sharply peaked to reproduce the observed broad feature (see the right panel of Figure~\ref{fig:traj}). Among the wide solutions, the ``wide B$+$'' model fit is worse by {$\Delta \chi^2 \approx296$} with respect to ``wide A$-$,'' significantly increased by {$\approx245$} compared with light curve only.} As discussed in \S~\ref{subsec:vlti_observables}, such $u_0>0$ geometries are strongly disfavored by the constraints from multi-epoch VLTI data.

\begin{figure}[htbp]
	\centering
	\includegraphics[width=0.8\textwidth]{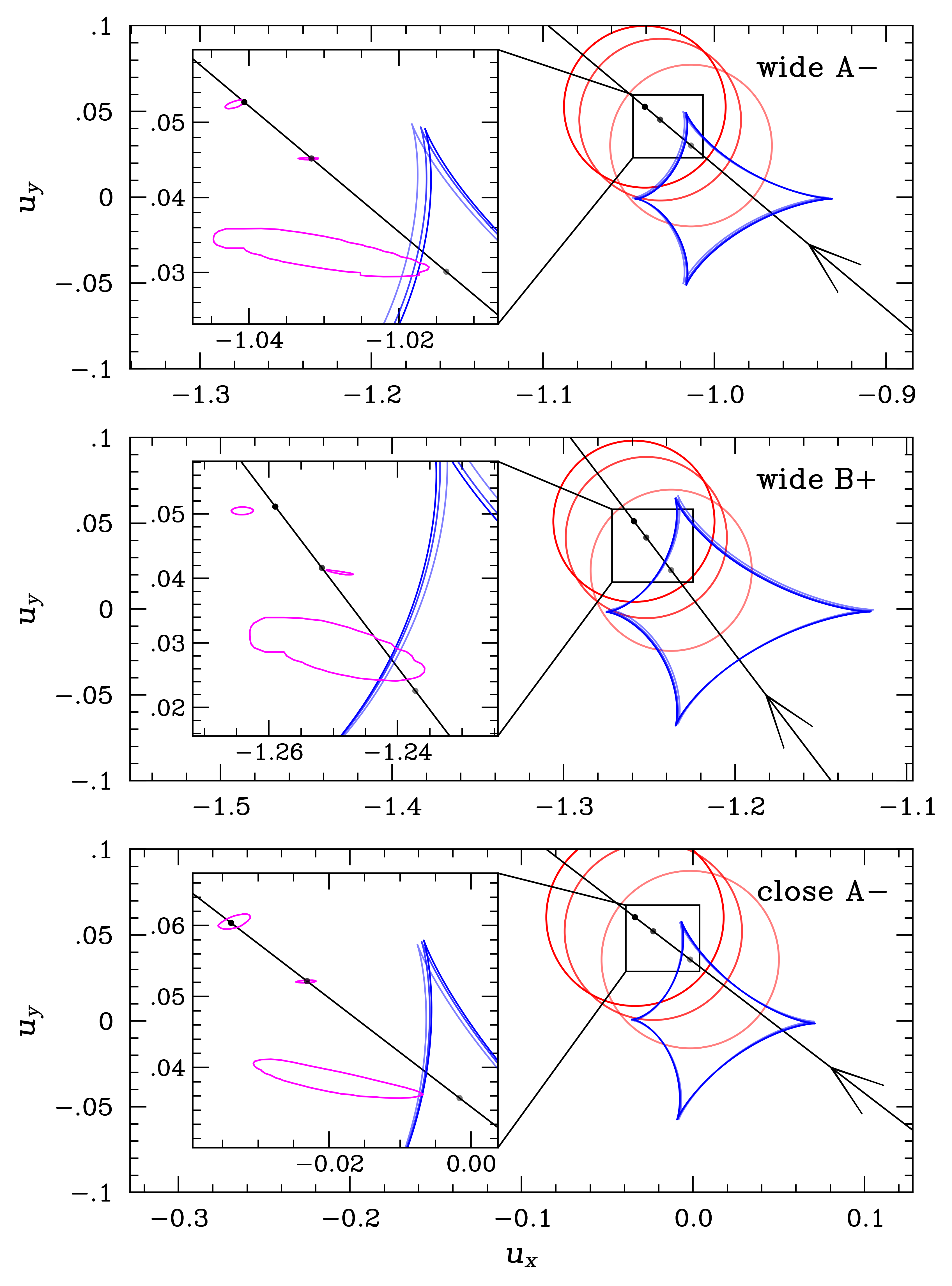}
	\caption{The source trajectory (black solid line) and caustics (blue closed curves) for \wide{} (top panel), ``wide B$+$'' (middle panel), and \close{} (bottom panel) solutions. The solid dots denote the source positions on the three nights of VLTI GRAVITY observations. The $3\sigma$ contours in magenta show the posteriors of the source positions inferred from the interferometric data only, with the geometric parameters $(s, q, \rho, \theta_\E, \Psi)$ held fixed to the values in Table~\ref{tab:joint_param}. }
	\label{fig:traj_consistency}
\end{figure}

To further understand the VLTI constraints on these three solutions, we performed additional fitting with the interferometric data only. We hold the parameters $(s, q, \rho, \Psi, \theta_\E)$, which define the caustics' configuration, fixed at their best-fit values from the joint modeling, while the source positions $(u_x, u_y)$ of the three nights are allowed to vary. Figure~\ref{fig:traj_consistency} shows the $3\sigma$ $(u_x, u_y)$ contours (magenta) compared with the expected source positions (black dots) on the best-fit trajectory (black solid line) from the joint modeling for each solution. For the \wide{} solution, there is a $\gtrsim3\sigma$ discrepancy in $(u_x, u_y)$ during Night 1, which corresponds to the $\lesssim0.\deg3$ systematic residuals of the closure phases for U4U2U1 and U4U3U2 in both exposures of that night, as displayed in the top two subpanels of Figure~\ref{fig:cpfit}. During the other two nights, the source positions are in good agreement with the \wide{} solution. Note that such $\sim0.3 \deg$ systematic offsets also show up in the third exposure of Night 2 (see Figure~\ref{fig:cpfit}), but they are not present in the other three exposures (i.e., Night 2-1, 2-2, and 2-4), which are in good agreement with the model. This suggests the existence of sub-degree-level systematic error in the closure-phase data. The \close{} solution shows a similar degree of consistency to the ``wide A$-$.'' In contrast, for the ``wide B+'' solution, which is ruled out by the VLTI data, the source positions of all three nights deviate by $>3\sigma$ from the expectation.

In conclusion, because the \wide{} solution is preferred over all other local minima by {$\Delta \chi^2\gtrsim 45$}, we regard it as the only favored solution of ASASSN-22av.

\section{Physical Parameters}\label{subsec:phy_param}
\subsection{Source Properties and Finite-source Effects}
\label{subsec:source_param}
The finite-source effects exhibited in ASASSN-22av's light curves provide an opportunity to cross-check the VLTI-measured $\theta_\E$ with $\theta _\E = {\theta_*}/{\rho}$. The standard method \citep{Yoo2004} for deriving $\theta_*$ for bulge microlensing events uses the source's dereddened color and magnitude with the extinction coefficients estimated from the location of bulge red clump (RC) stars in a color-magnitude diagram (CMD). In our case, the source star is not toward the bulge, and the CMD contains RC stars at a variety of distances, making it difficult to apply the standard method.
	{Gaia} DR3 reports a trigonometric parallax $\pi_\Sc = 0.137 \pm 0.016\,\mas$. We adopt $\pi_{\Sc, \rm corr} = 0.185\,\mas$ after applying the preliminary zero-point correction from \citet{parallax_zero_point}. The independent investigation on the {Gaia} parallax zero-point offset by \citet{parallax_zero_point2} yields a consistent value of $\pi _{\Sc, \rm corr} = 0.207\pm 0.025\,\mas$.
We fit the spectral energy distribution (SED) using photometric data from 2MASS, {Gaia} and SkyMapper Southern Survey DR4 \citep{smssdr4}. We employ the MESA Isochrones and Stellar Tracks \citep[MIST;][]{MIST} to generate synthetic magnitudes as a function of stellar mass, age, and metallicity.
The extinction $E(B-V)$ is taken as a free parameter with a flat prior, and we assume that $R_V = 3.1$. We use the corrected {Gaia} parallax as a prior for source distance.
MIST returns the stellar radius $R_*$, and hence the angular radius $\theta_* = R_* / D_\Sc$ can be estimated with MCMC. The photometric error bars are inflated by a factor of 2, so that $\chi^2/$dof $\sim1$. We find that the best-fit $E(B-V)=0.64^{+0.06}_{-0.04}$, which is consistent with the lower limit of $E(B-V)> 0.61$ (for distance greater than $4.3$\,kpc), according to the three-dimensional dust map of \citet{Guo2021}.
The best-fit stellar parameters are $T_\eff = 3747^{+46}_{-18}\,{\rm K} ,\,  \log g = 1.07^{+0.11}_{-0.04}$ and $[{\rm Fe/H}] = 0.01 ^{+0.12}_{-0.07}$, arriving at $\theta_* = 44.0 \pm 0.8\,\mu \rm as$. The SED-derived stellar parameters are consistent with the spectroscopic ones from \S~\ref{sec:spectrum}.

For the best-fit \wide{} solution with {$\rho = 0.058\pm 0.001$}, we find {$\theta _\E = 0.759 \pm 0.019\,\mas$}, which differs by {$1.8\,\sigma$} from the {$\theta _\E = 0.724\pm0.002\,\mas$} determined by VLTI GRAVITY.
In comparison, for the \close{} solution, where {$\rho = 0.051^{+0.001}_{-0.001}$}, the corresponding {$\theta_\E = 0.862 \pm 0.023\,\mas$} deviates from the VLTI GRAVITY value of {$\theta _\E \simeq 0.737\pm0.002\,\mas$} by {$5.4 \sigma$}. Note that the $\sim2\%$ error of our $\theta_*$ estimate is likely an underestimate, due to unaccounted-for systematic uncertainties. For example, $\theta_*$ inferences from the empirical color/surface brightness relations such as \citet{kervella04} generally have errors of $\sim 4\%$\textendash$5\%$. In any case, for the \wide{} solution, $\theta_\E$ derived from VLTI and finite-source effects are {basically consistent}.

\subsection{Physical Properties of the Lens System}
\label{subsec:lens_param}
With measurements of the angular Einstein radius $\theta_\E $ and the microlens parallax $\pi_\E$, the physical parameters of the lens system can be unambiguously derived from microlensing observables:
\begin{equation}
	M_\L = \frac{\theta_\E}{\kappa \pi_\E}; \quad \pi_\rel = \theta_\E\pi_\E; \quad \bmu_\rel = \frac{\theta_\E}{ t_\E}\,\frac{\bpi_\E}{\pi_\E},
\end{equation}
and adopting $\pi _\Sc = 0.185\pm0.016 \,\mas$, we can also infer the lens distance:
\begin{equation}
	D_\L = \frac{\au}{\pi_\E \theta _\E + \pi _\Sc}.
\end{equation}
The lens properties of the favored solution (\wide{}) are listed in Table~\ref{tab:phy_param}. The lens system is composed of two M dwarfs with {$M_1= 0.256^{+0.008}_{-0.006}\, M_\odot$ and $M_2 = 0.130\pm{0.007}\, M_\odot$, separated by a projected separation of $r_\perp = 6.83\pm0.31\,\au$ and at a distance of $D_\L=2.29\pm0.08$\,kpc.}

Based on MIST isochrones, we estimate that the lens system has $K_s=18.5$ (assuming zero extinction). In comparison, the source's 2MASS magnitude is $K_s = 9.27$. During the VLTI observations, the magnification is more than $15$, and thus the flux ratio between the lens and magnified source is $\xi= f_{\L} / (Af_{\Sc}) < 1.4 \times 10^{-5}$.  Therefore, the lens light has a negligible effect on the closure phase, justifying our approach of calculating closure phases without the lens flux contribution.

\section{Discussion and Conclusion}

ASASSN-22av is the first binary-lens microlensing event with multiple images resolved by interferometry. The modeling is considerably more complex than for a single-lens event. For a typical single-lens event, the simple model of two point images is generally satisfactory for fitting the closure-phase data \citep{Dong2019}, and thus closure-phase modeling can, in principle, be performed separately from the light curve. In contrast, a binary lens has three or five images for a point source, depending on whether the source is outside or inside the caustics. During the VLTI observations of ASASSN-22av, the images, formed from a finite source crossing caustics, morphed from a nearly complete Einstein ring to separated arcs over a period of several nights. Thus, the interpretation of the closure-phase data from such a binary-lens event requires inputs from light-curve modeling.

The weak binary-lens perturbation and sparse pre-peak photometric coverage make the light-curve modeling particularly complicated for ASASSN-22av. In the early stages of our work, where we did not include the {Gaia} light curve, we found 25 local minima in our binary-lens grid search. After incorporating the VLTI data, the best-fit solution, which is the same as the presently favored \wide{} solution, `predicted' the existence of the bump at $\rm HJD' \sim 9300$ shown in the {Gaia} light curve. We only realized this `prediction' later, after conducting light-curve modeling after the inclusion of {Gaia} data. The predictive power of the \wide{} solution offers good evidence for its validity.

We also made several consistency checks between the light curve and multi-epoch VLTI data. For the favored solution, there are good agreements in multiple aspects, including $\alpha$ (Figure~\ref{fig:alpha}), direction of $\bpi_\E$ (Figure~\ref{fig:piE}), source trajectory (Figure~\ref{fig:traj_consistency}), and $\theta_\E$ (see the discussion in \S~\ref{subsec:source_param}), further demonstrating the model's robustness.
The agreement between the best-fit model and the closure-phase data is at $\lesssim 0.3\deg$ (Figure~\ref{fig:cpfit}, \S \ref{subsec:joint_analysis}), which likely reflects the instrument's systematics floor. We measure the Einstein radius {$\theta_\E = 0.724$\,mas with $0.3\%$ precision}, even though it is close to the smallest value resolvable by VLTI GRAVITY. The expected closure phase would be $\lesssim 5 \deg$ for $\theta_\E$ to $0.6\,\mas$ and $\lesssim 2\deg$ for $\theta_\E = 0.5\,\mas$.

The lens system consists of two M dwarfs separated by {$\sim 6.8\,\au$ at $2.3\,\rm kpc$, with $(M_1, M_2)=(0.256\pm{0.008}, 0.130\pm{0.007})\,  M_\sun$. The primary mass is more precisely determined than the secondary ($3.1\%$ versus $5.4\%$).} This is because, in the wide-binary ($s>1$) regime, the scale of images (measured by VLTI GRAVITY) is proportional to $\theta_\E$ of the primary mass rather than the total mass. Hence, the uncertainty of the secondary mass $M_2$ is contributed by an additional uncertainty from the mass ratio.
The estimated orbital period is {$P\sim 28\,\rm yr$} for a circular orbit, and the angular separation between the binaries is $\sim 3\,\mas$ at $2.3 \,\rm kpc$. Such faint low-mass binary systems at intermediate orbital separations have so far been inaccessible by any other method \citep[see, e.g., Figure~1 of ][]{binarydetection}. Similar VLTI microlensing observations will be able to identify and characterize intermediate binaries containing dark compact objects like neutron stars and stellar-mass black holes to complement other detection techniques.

\section*{Acknowledgment}
We thank Cheongho Han and Gregory Green for helpful discussions. This work is based on observations collected at the European Southern Observatory under Director Discretionary Time program 108.23MK. This work is supported by the National Natural Science Foundation of China (grant No. 12133005) and the science research grants from the China Manned Space Project with No. CMS-CSST-2021-B12. S.D. acknowledges the New Cornerstone Science Foundation through the XPLORER PRIZE. C.S.K. is supported by NSF grants AST-2307385 and 2407206. This research was funded in part by the National Science Centre, Poland, and grants OPUS 2021/41/B/ST9/00252 and SONATA 2023/51/D/ST9/00187 awarded to P.M.

This research uses data obtained through the Telescope Access Program (TAP), which has been funded by the TAP member institutes. This work has made use of data from the European Space Agency (ESA) mission {Gaia} (\url{https://www.cosmos.esa.int/gaia}), processed by the {Gaia} Data Processing and Analysis Consortium (DPAC;
\url{https://www.cosmos.esa.int/web/gaia/dpac/consortium}). Funding for the DPAC has been provided by national institutions, in particular the institutions participating in the {Gaia} Multilateral Agreement. We acknowledge the Gaia Photometric Science Alerts Team. We also acknowledge the devoted staff of the Anglo-Australian Telescope who have maintained the HERMES spectrograph over a decade.

This research has made use of the VizieR catalog access tool, CDS, Strasbourg, France. This publication makes use of data products from the Two Micron All Sky Survey, which is a joint project of the University of Massachusetts and the Infrared Processing and Analysis Center/California Institute of Technology, funded by the National Aeronautics and Space Administration and the National Science Foundation.

\appendix
\section{Binning the closure-phase data}
\label{appendix:binning}
The interferometric observables are the complex visibilities $\mathcal{V}_{pq}$ and closure phases $\phi_{pqr}$, with
\citep{vanCittert1934, Zernike1938}
\begin{equation}
	\mathcal{V}_{pq}(u, v; \lambda ) = \frac{\int I(\alpha ,\delta )e^{-2\pi i(u\alpha +v\delta )}d\alpha d\delta }{\int I(\alpha ,\delta )d\alpha d\delta },
\end{equation}
where $I(\alpha, \delta)$ is the on-sky target intensity distribution centered on the light of sight $(\alpha_0, \delta_0)$ and $(u, v) = (\bm n_{\rm E}, \bm n_{\rm N})\cdot\bm B_{pq} / \lambda$ is the on-sky projection of the baseline vector over wavelength.
The closure phase is defined for a three-telescope array as $\phi_{pqr}(\lambda) = \arg (\mathcal{V}_{pq} \mathcal{V}_{qr} \mathcal{V}_{rp})$. For a list of closure phases $\{\phi_k\}$, the mean closure phase $\bar \phi$ is calculated by assigning $\phi _k$ to the unit complex and summing them up for binning as
\begin{equation}
	\bar\phi = \arg \sum_k e^{i\phi_k}.
\end{equation}
The uncertainty of $\bar \phi $ is estimated using bootstrap resampling.

\section{Image generation and visibility calculation}
\label{appendix:binaryimage}
For a given set of source coordinate $(u_x, u_y)$ and binary configuration $(s,q)$, we extract the image contours used for the magnification calculation in \texttt{VBBinaryLensing} package. The contours are then converted to two-dimensional images using the loop-linking method \citep{Dong2006}, which combines the contour integration and inverse-ray-shooting methods.
The loop-linking method fills the contour with a uniform grid of points organized in links.
	{At a given step size $\delta$, our code returns the head $(x_k, y_k)$ and the number of points $n_{k}$ of the $k$th link, which contains a set of points $(x_k, y_k + j\delta)$ with $j \in \{0, 1, 2\cdots n_k - 1\}$}. For any baseline coordinate $(u, v)$, the complex visibility is calculated by summing up the contribution from all links (assuming $\Psi=0$) as
\begin{align*}
	\mathcal{V}(u, v) & = \sum_k \sum_{j=0}^{n_k - 1} e^{-2\pi i (v x_k + u (y_k + j\delta ))}                              \\
	                  & = \sum_k  e^{-2\pi i (v x_k + u y_k)} \sum_{j=0}^{n_k - 1} (e^{-2\pi i u\delta})^j                  \\
	                  & = \sum_k  e^{-2\pi i (v x_k + u y_k)} \frac{1 - e^{-2\pi i u\delta n_k}}{1 - e^{-2\pi i u\delta }}.
\end{align*}
The calculation is optimized by grouping the links by $n_k$ and summing up them first, which reduces the number of calculations by a factor of $\sim 2$.
We caution that the dot product of $(x_k, y_k)$ and $(u, v)$ is not arbitrary. In our case, it has to be $x_k v + y_k u$. Otherwise, the transformation from the image plane to the sky plane is not positive-definite and the angle $\Psi$ is not well-defined.

An alternative method for computing the visibility is applying the divergence theorem on the closed image contour $\partial D$ with
\begin{equation}
	V(\bm u) = \iint_{D} e^{-2\pi i \bm u \cdot \bm r} d^2 \bm r = \oint_{\partial D} e^{-2\pi i \bm u \cdot \bm r} \frac{\bm u \cdot \bm {\hat n} }{-2\pi i u ^2 } ds,
\end{equation}
where $\bm {\hat n}$ is the unit vector normal to the boundary $\partial D$ at $\bm r$.
For a closed polygon with vertices $\bm r_k$, the integration for edge $\bm r_k \to \bm r_{k+1}$ is
\begin{equation}
	\int_{\bm r_k}^{\bm r_{k+1}} e^{-2\pi i \bm u \cdot \bm r} \frac{\bm u \times d\bm {r} }{-2\pi i u ^2 } =
	\frac{1}{4\pi ^2 u^2 }\frac{\bm u \times (\bm r_k - \bm r_{k+1})}{\bm u \cdot (\bm r_k - \bm r_{k+1})} (e^{-2\pi i \bm u \cdot \bm r_{k}} - e^{-2\pi i \bm u \cdot \bm r_{k+1}}).
\end{equation}
The visibility is then the sum over all edges. This method is faster than the loop-linking method for convex images, while the loop-linking method is faster for concave images. In our case, the images are ring-like arcs, and thus we adopt the loop-linking method.
\software{{PmPyeasy}, {hotpants}, {DoPHOT}, {VBBinaryLensing}, {EMCEE}, {PMOIRED}, {MIST}}

\input{ref.tex}
\input{orb.tex}
\input{joint.tex}
\input{phy.tex}

\end{document}

%% file: orb.tex
\begin{table*}[htbp]
	\centering
	\caption{Best-fit light-curve parameters and $1\sigma $ uncertainties}
	\label{tab:lc_param}
	\begin{tabular}{lcccccccccccc}
		\hline \hline
		Solution & $t_0$              & $u_0$                & $t_\mathrm{E}$   & $\rho$               & $s$              & $q$                & $\alpha$         & $\pi_{\rm E,N}$    & $\pi_{\rm E,E}$    & $ds/dt$          & $d \alpha/dt$         & $\beta $           \\
		$\chi^2$   & $(\rm HJD')$         &                      & $(\rm days)$        &                      &                  &                    & $(\rm deg)$        &                    &                    & $(\rm yr ^{-1})$  & $(\rm yr ^{-1})$ &                    \\
		\hline
close A$+$ & $9601.24$ & $0.035$ & $65.4$ & $0.056$ & $0.4$ & $0.36$ & $37.4$ & $-0.28$ & $-0.14$ & $0.15$ & $-1.41$ & $0.03$ \\
1005.3 & $^{+0.04}_{-0.05}$ & $^{+0.002}_{-0.002}$ & $^{+1.3}_{-1.1}$ & $^{+0.002}_{-0.002}$ & $^{+0.1}_{-0.1}$ & $^{+0.04}_{-0.03}$ & $^{+0.6}_{-0.6}$ & $^{+0.04}_{-0.03}$ & $^{+0.02}_{-0.02}$ & $^{+0.02}_{-0.02}$ & $^{+0.07}_{-0.06}$ & $^{+0.01}_{-0.01}$ \\
close A$-$ & $9601.11$ & $-0.028$ & $74.0$ & $0.048$ & $0.4$ & $0.21$ & $-38.1$ & $-0.14$ & $-0.10$ & $0.20$ & $2.11$ & $0.12$ \\
1013.5 & $^{+0.06}_{-0.04}$ & $^{+0.002}_{-0.002}$ & $^{+1.4}_{-1.8}$ & $^{+0.003}_{-0.002}$ & $^{+0.1}_{-0.1}$ & $^{+0.04}_{-0.02}$ & $^{+0.6}_{-0.6}$ & $^{+0.03}_{-0.03}$ & $^{+0.02}_{-0.02}$ & $^{+0.02}_{-0.03}$ & $^{+0.05}_{-0.08}$ & $^{+0.02}_{-0.03}$ \\
close B$+$ & $9601.35$ & $0.040$ & $63.5$ & $0.056$ & $0.4$ & $0.50$ & $-47.9$ & $-0.27$ & $-0.09$ & $0.16$ & $1.49$ & $0.03$ \\
1025.9 & $^{+0.03}_{-0.03}$ & $^{+0.002}_{-0.002}$ & $^{+0.9}_{-1.1}$ & $^{+0.002}_{-0.002}$ & $^{+0.1}_{-0.1}$ & $^{+0.05}_{-0.05}$ & $^{+0.5}_{-0.5}$ & $^{+0.03}_{-0.02}$ &
$^{+0.02}_{-0.02}$ & $^{+0.01}_{-0.01}$ & $^{+0.06}_{-0.06}$ & $^{+0.01}_{-0.01}$ \\
\hline
wide A$+$ & $9601.34$ & $0.035$ & $56.3$ & $0.068$ & $5.4$ & $1.04$ & $33.2$ & $-0.40$ & $-0.03$ & $0.80$ & $0.16$ & $0.98$ \\
1028.4 & $^{+0.03}_{-0.04}$ & $^{+0.002}_{-0.002}$ & $^{+1.4}_{-1.3}$ & $^{+0.002}_{-0.003}$ & $^{+0.2}_{-0.2}$ & $^{+0.08}_{-0.07}$ & $^{+0.6}_{-0.5}$ & $^{+0.03}_{-0.03}$ & $^{+0.02}_{-0.02}$ & $^{+0.14}_{-0.15}$ & $^{+0.03}_{-0.04}$ & $^{+0.02}_{-0.05}$ \\
\bf wide A$-$ & $\bf 9601.30$ & $\bf -0.028$ & $\bf 69.5$ & $\bf 0.054$ & $\bf 5.1$ & $\bf 0.77$ & $\bf -38.5$ & $\bf -0.13$ & $\bf -0.24$ & $\bf 0.19$ & $\bf -0.11$ & $\bf 0.39$ \\
\bf 948.9 & $^{+0.08}_{-0.07}$ & $^{+0.002}_{-0.002}$ & $^{+3.0}_{-3.3}$ & $^{+0.002}_{-0.002}$ & $^{+0.3}_{-0.3}$ & $^{+0.09}_{-0.08}$ & $^{+1.1}_{-0.9}$ & $^{+0.06}_{-0.08}$ & $^{+0.02}_{-0.01}$ & $^{+0.34}_{-0.33}$ & $^{+0.06}_{-0.05}$ & $^{+0.31}_{-0.18}$ \\
wide B$+$ & $9601.07$ & $0.033$ & $61.1$ & $0.054$ & $6.2$ & $1.27$ & $-39.0$ & $-0.22$ & $-0.26$ & $1.03$ & $-0.00$ & $0.87$ \\
999.6 & $^{+0.04}_{-0.04}$ & $^{+0.002}_{-0.002}$ & $^{+1.9}_{-1.9}$ & $^{+0.003}_{-0.002}$ & $^{+0.3}_{-0.3}$ & $^{+0.12}_{-0.09}$ & $^{+1.1}_{-1.1}$ & $^{+0.07}_{-0.07}$ & $^{+0.01}_{-0.01}$ & $^{+0.07}_{-0.14}$ & $^{+0.04}_{-0.03}$ & $^{+0.07}_{-0.20}$ \\
		\hline
	\end{tabular}
    \tablecomments{For close solutions ($s<1$), $t_0$ and $u_0$ are defined relative to the barycenter of the lens system, and the parameters $\{t_\E, s, \rho, \pi_{\rm E,N}, \pi_{\rm E,E}\}$ are in units of the $\theta _\E$ of the total mass. For wide solutions ($s>1$), $t_0$ and $u_0$ are defined relative to the magnification center of the primary lens \citep{Dong2006}, and the parameters $\{t_\E, s, \rho, \pi_{\rm E,N}, \pi_{\rm E,E}\}$ are defined in units of the $\theta _\E$ of the primary lens. The favored solution \wide{} is in boldface.}
\end{table*}

%% file: joint.tex
\begin{table*}[htbp]
	\centering
	\caption{Best-fit parameters and $1\sigma $ uncertainties from the joint VLTI and light-curve analysis}
	\label{tab:joint_param}
	\begin{tabular}{lccccccccccccc}
		\hline \hline
		Solution   & $t_0$              & $u_0$                & $t_\mathrm{E}$   & $\rho$               & $s$              & $q$                & $\alpha$         & $\pi_{\rm E}$      & $ds/dt$            & $d \alpha/dt$      & $\beta $           & $\theta_\E$          & $\Psi$           \\
		$\chi^2$   & $(\rm HJD')$         &                      & $(\rm days)$        &                      &                  &                    & $(\rm deg)$        &                    & $(\rm yr ^{-1} )$    & $(\rm yr ^{-1} )$    &                    & $(\mas)$               & $(\rm deg)$        \\
		\hline

		close A$-$ & $9601.14$          & $-0.027$             & $74.4$           & $0.051$              & $0.4$            & $0.23$             & $-37.4$          & $0.18$             & $0.18$             & $2.09$             & $0.10$             & $0.737$              & $258.6$          \\
		1220.0     & $^{+0.02}_{-0.02}$ & $^{+0.001}_{-0.001}$ & $^{+0.7}_{-0.6}$ & $^{+0.001}_{-0.001}$ & $^{+0.1}_{-0.1}$ & $^{+0.01}_{-0.01}$ & $^{+0.4}_{-0.4}$ & $^{+0.01}_{-0.01}$ & $^{+0.01}_{-0.01}$ & $^{+0.02}_{-0.03}$ & $^{+0.01}_{-0.01}$ & $^{+0.002}_{-0.002}$ & $^{+0.2}_{-0.2}$ \\
		\bf wide A$-$  & $\bf9601.12$          & $\bf-0.028$             & $\bf70.1$           & $\bf0.058$              & $\bf4.1$            & $\bf0.51$             & $\bf-40.0$          & $\bf0.35$             & $\bf-0.78$            & $\bf-0.09$            & $\bf0.47$             & $\bf0.724$              & $\bf259.4$          \\
		1174.8     & $^{+0.04}_{-0.04}$ & $^{+0.001}_{-0.001}$ & $^{+1.1}_{-0.9}$ & $^{+0.001}_{-0.001}$ & $^{+0.1}_{-0.1}$ & $^{+0.03}_{-0.03}$ & $^{+0.5}_{-0.5}$ & $^{+0.02}_{-0.01}$ & $^{+0.20}_{-0.18}$ & $^{+0.03}_{-0.03}$ & $^{+0.23}_{-0.17}$ & $^{+0.002}_{-0.002}$ & $^{+0.2}_{-0.2}$ \\
		wide B$+$  & $9600.68$          & $-0.005$             & $63.9$           & $0.062$              & $4.2$            & $0.76$             & $-53.0$          & $0.46$             & $0.49$             & $-0.14$            & $0.26$             & $0.726$              & $262.5$          \\
		1470.5     & $^{+0.03}_{-0.03}$ & $^{+0.001}_{-0.001}$ & $^{+1.1}_{-1.2}$ & $^{+0.001}_{-0.001}$ & $^{+0.1}_{-0.1}$ & $^{+0.01}_{-0.01}$ & $^{+0.5}_{-0.6}$ & $^{+0.02}_{-0.02}$ & $^{+0.20}_{-0.20}$ & $^{+0.04}_{-0.04}$ & $^{+0.07}_{-0.06}$ & $^{+0.002}_{-0.002}$ & $^{+0.2}_{-0.2}$ \\
		\hline
	\end{tabular}
	\tablecomments{The light-curve parameters are defined in the same fashion as in  Table~\ref{tab:lc_param}. For wide solutions, $\theta_\E$ is defined by the mass of the primary lens, while for close solutions, it is defined by the total mass of the lens system. The favored solution \wide{} is in boldface.}
\end{table*}

%% file: phy.tex
\begin{table*}[htbp]
	\caption{Derived physical parameters of the lens system}
	\label{tab:phy_param}
	\centering
	\begin{tabular}{ccccccc}
		\hline \hline
		$M_1$                     & $M_2$                     & $r_\perp$              & $\pi_{\rm rel}$           & $D_\L$                 & $\mu_{\rm hel,N}$       & $\mu_{\rm hel, E}$      \\
		$(M_\sun)$                  & $(M_\sun)$                  & $(\au)$                  & $(\mas)$                    & $(\rm kpc)$              & $(\rm mas\,yr ^{-1})$     & $(\rm mas\, yr ^{-1})$    \\
		\hline
		$0.256^{+0.008}_{-0.006}$ & $0.130^{+0.007}_{-0.007}$ & $6.83^{+0.31}_{-0.27}$ & $0.251^{+0.007}_{-0.007}$ & $2.29^{+0.08}_{-0.08}$ & $-1.36^{+0.05}_{-0.05}$ & $-2.66^{+0.06}_{-0.05}$ \\
		\hline
	\end{tabular}
\end{table*}

%% file: ms.bbl
\begin{thebibliography}{99}
    \bibitem[Afonso et al.(2001)]{Afonso2001} Afonso, C., Albert, J.~N., Andersen, J., et al.\ 2001, \aap, 378, 1014. doi:10.1051/0004-6361:20011204
    \bibitem[Alard \& Lupton(1998)]{alard_luption1998} Alard, C. \& Lupton, R.~H.\ 1998, \apj, 503, 325. doi:10.1086/305984
    \bibitem[Alard(2000)]{alard_image_sub2000} Alard, C.\ 2000, \aaps, 144, 363. doi:10.1051/aas:2000214
    \bibitem[Albrow et al.(2001)]{Albrow2001} Albrow, M., An, J., Beaulieu, J.-P., et al.\ 2001, \apjl, 550, L173. doi:10.1086/319635
    \bibitem[An et al.(2002)]{An2002} An, J.~H., Albrow, M.~D., Beaulieu, J.-P., et al.\ 2002, \apj, 572, 521. doi:10.1086/340191
	\bibitem[Batista et al.(2011)]{Batista2011} Batista, V., Gould, A., Dieters, S., et al.\ 2011, \aap, 529, A102. doi:10.1051/0004-6361/201016111
	\bibitem[Becker(2015)]{Becker2015} Becker, A.\ 2015, Astrophysics Source Code Library. ascl:1504.004
	\bibitem[Bernstein et al.(2003)]{MIKE} Bernstein, R., Shectman, S.~A., Gunnels, S.~M., et al.\ 2003, \procspie, 4841, 1694. doi:10.1117/12.461502
	\bibitem[Bozza(2010)]{VBBL2010} Bozza, V.\ 2010, \mnras, 408, 2188. doi:10.1111/j.1365-2966.2010.17265.x
	\bibitem[Bozza et al.(2018)]{VBBL2018} Bozza, V., Bachelet, E., Bartoli{\'c}, F., et al.\ 2018, \mnras, 479, 5157. doi:10.1093/mnras/sty1791
	\bibitem[Brown et al.(2013)]{Brown2013} Brown, T.~M., Baliber, N., Bianco, F.~B., et al.\ 2013, \pasp, 125, 1031. doi:10.1086/673168
	\bibitem[Buckley et al.(2006)]{SALT} Buckley, D.~A.~H., Swart, G.~P., \& Meiring, J.~G.\ 2006, \procspie, 6267, 62670Z. doi:10.1117/12.673750
    \bibitem[Buder et al.(2018)]{Buder2018} Buder, S., Asplund, M., Duong, L., et al.\ 2018, \mnras, 478, 4513. doi:10.1093/mnras/sty1281
    \bibitem[Buder et al.(2019)]{Buder2019} Buder, S., Lind, K., Ness, M.~K., et al.\ 2019, \aap, 624, A19. doi:10.1051/0004-6361/201833218
    \bibitem[Buder et al.(2021)]{Buder2021} Buder, S., Sharma, S., Kos, J., et al.\ 2021, \mnras, 506, 150. doi:10.1093/mnras/stab1242
    \bibitem[Buder et al.(2024)]{Buder2024} Buder, S., Kos, J., Wang, E.~X., et al.\ 2024, arXiv:2409.19858. doi:10.48550/arXiv.2409.19858
	\bibitem[Cassan \& Ranc(2016)]{vltimicrolens4} Cassan, A., \& Ranc, C.\ 2016, \mnras, 458, 2074
    \bibitem[Cassan et al.(2022)]{Cassan2022} Cassan, A., Ranc, C., Absil, O., et al.\ 2022, Nature Astronomy, 6, 121. doi:10.1038/s41550-021-01514-w
	\bibitem[Chang \& Refsdal(1979)]{ChangRefsdal1979} Chang, K. \& Refsdal, S.\ 1979, \nat, 282, 561. doi:10.1038/282561a0
	\bibitem[Chen et al.(2022)]{Chen2022} Chen, P., Dong, S., Kochanek, C.~S., et al.\ 2022, \apjs, 259, 53. doi:10.3847/1538-4365/ac50b7
    \bibitem[Choi et al.(2016)]{MIST} Choi, J., Dotter, A., Conroy, C., et al.\ 2016, \apj, 823, 102. doi:10.3847/0004-637X/823/2/102
	\bibitem[Claret \& Bloemen(2011)]{Claret2011} Claret, A. \& Bloemen, S.\ 2011, \aap, 529, A75. doi:10.1051/0004-6361/201116451
    \bibitem[Claret(2019)]{Claret2019} Claret, A.\ 2019, VizieR Online Data Catalog, 6154. VI/154
	\bibitem[Dalal \& Lane(2003)]{vltimicrolens2} Dalal, N., \& Lane, B.~F.\ 2003, \apj, 589, 199
	\bibitem[Delplancke et al.(2001)]{vltimicrolens1} Delplancke, F., G{\'o}rski, K.~M., \& Richichi, A.\ 2001, \aap, 375, 701
    \bibitem[De Silva et al.(2015)]{DeSilva2015} De Silva, G.~M., Freeman, K.~C., Bland-Hawthorn, J., et al.\ 2015, \mnras, 449, 2604. doi:10.1093/mnras/stv327
	\bibitem[Dong et al.(2006)]{Dong2006} Dong, S., DePoy, D.~L., Gaudi, B.~S., et al.\ 2006, \apj, 642, 842. doi:10.1086/501224
	\bibitem[Dong et al.(2007)]{Dong2007} Dong, S., Udalski, A., Gould, A., et al.\ 2007, \apj, 664, 862. doi:10.1086/518536
	\bibitem[Dong et al.(2009)]{Dong2009} Dong, S., Gould, A., Udalski, A., et al.\ 2009, \apj, 695, 970. doi:10.1088/0004-637X/695/2/970
    \bibitem[Dong et al.(2019)]{Dong2019} Dong, S., M{\'e}rand, A., Delplancke-Str{\"o}bele, F., et al.\ 2019, \apj, 871, 70. doi:10.3847/1538-4357/aaeffb
    \bibitem[Einstein(1936)]{Einstein1936} Einstein, A.\ 1936, Science, 84, 506. doi:10.1126/science.84.2188.506
    \bibitem[Eisenhauer et al.(2023)]{Eisenhauer2023} Eisenhauer, F., Monnier, J.~D., \& Pfuhl, O.\ 2023, \araa, 61, 237. doi:10.1146/annurev-astro-121622-045019
    \bibitem[El-Badry(2024)]{binarydetection} El-Badry, K.\ 2024, \nar, 98, 101694. doi:10.1016/j.newar.2024.101694
	\bibitem[Foreman-Mackey et al.(2013)]{Foreman-Mackey2013} Foreman-Mackey, D., Hogg, D.~W., Lang, D., et al.\ 2013, \pasp, 125, 306. doi:10.1086/670067
	\bibitem[Gaia Collaboration et al.(2016)]{Gaia2016} Gaia Collaboration, Prusti, T., de Bruijne, J.~H.~J., et al.\ 2016, \aap, 595, A1. doi:10.1051/0004-6361/201629272
	\bibitem[Gaia Collaboration et al.(2023)]{GaiaDR3} Gaia Collaboration, Vallenari, A., Brown, A.~G.~A., et al.\ 2023, \aap, 674, A1. doi:10.1051/0004-6361/202243940
 	\bibitem[Gould(1992)]{Gould1992} Gould, A.\ 1992, \apj, 392, 442. doi:10.1086/171443
	\bibitem[Gould(1994a)]{Gould1994a} Gould, A.\ 1994, \apjl, 421, L71. doi:10.1086/187190
	\bibitem[Gould(1994b)]{Gould1994b} Gould, A.\ 1994, \apjl, 421, L75. doi:10.1086/187191
	\bibitem[Gould et al.(1994)]{Gould1d1994} Gould, A., Miralda-Escude, J., \& Bahcall, J.~N.\ 1994, \apjl, 423, L105. doi:10.1086/187247
	\bibitem[Gould(2000)]{Gould2000} Gould, A. 2000, \apj, 542, 785
	\bibitem[Gould(2004)]{Gould2004} Gould, A.\ 2004, \apj, 606, 319. doi:10.1086/382782
    \bibitem[GRAVITY Collaboration et al.(2017)]{gravity2017} GRAVITY Collaboration, Abuter, R., Accardo, M., et al.\ 2017, \aap, 602, A94. doi:10.1051/0004-6361/201730838
    \bibitem[GRAVITY+ Collaboration et al.(2022a)]{gravitywide} GRAVITY+ Collaboration, Abuter, R., Allouche, F., et al.\ 2022a, \aap, 665, A75. doi:10.1051/0004-6361/202243941
    \bibitem[GRAVITY+ Collaboration et al.(2022b)]{gravityplus} GRAVITY+ Collaboration, Abuter, R., Alarcon, P., et al.\ 2022b, The Messenger, 189, 17. doi:10.18727/0722-6691/5285
   \bibitem[Groenewegen(2021)]{parallax_zero_point2} Groenewegen, M.~A.~T.\ 2021, \aap, 654, A20. doi:10.1051/0004-6361/202140862
	\bibitem[Guo et al.(2021)]{Guo2021} Guo, H.-L., Chen, B.-Q., Yuan, H.-B., et al.\ 2021, \apj, 906, 47. doi:10.3847/1538-4357/abc68a
 	\bibitem[Hodgkin et al.(2021)]{Hodgkin2021} Hodgkin, S.~T., Harrison, D.~L., Breedt, E., et al.\ 2021, \aap, 652, A76. doi:10.1051/0004-6361/202140735
	\bibitem[Jung et al.(2022)]{Jung2022} Jung, Y.~K., Zang, W., Han, C., et al.\ 2022, \aj, 164, 262. doi:10.3847/1538-3881/ac9c5c
	\bibitem[Kammerer et al.(2020)]{Kammerer2020} Kammerer, J., M{\'e}rand, A., Ireland, M.~J., et al.\ 2020, \aap, 644, A110. doi:10.1051/0004-6361/202038563
	\bibitem[Kervella et al.(2004)]{kervella04} Kervella, P., Bersier, D., Mourard, D., et al.\ 2004, \aap, 428, 587
	\bibitem[Kochanek et al.(2017)]{asassn_v1} Kochanek, C.~S., Shappee, B.~J., Stanek, K.~Z., et al.\ 2017, \pasp, 129, 104502. doi:10.1088/1538-3873/aa80d9
    \bibitem[Kos et al.(2017)]{Kos2017} Kos, J., Lin, J., Zwitter, T., et al.\ 2017, \mnras, 464, 1259. doi:10.1093/mnras/stw2064
    \bibitem[Le Bouquin et al.(2011b)]{LeBouquin2011} Le Bouquin, J.-B., Berger, J.-P., Lazareff, B., et al.\ 2011, \aap, 535, A67. doi:10.1051/0004-6361/201117586
    \bibitem[Lindegren et al.(2021a)]{Lindegren2021a} Lindegren, L., Klioner, S.~A., Hern{\'a}ndez, J., et al.\ 2021a, \aap, 649, A2. doi:10.1051/0004-6361/202039709
    \bibitem[Lindegren et al.(2021b)]{parallax_zero_point} Lindegren, L., Bastian, U., Biermann, M., et al.\ 2021b, \aap, 649, A4. doi:10.1051/0004-6361/202039653
    \bibitem[Mayor et al.(2003)]{HARPS} Mayor, M., Pepe, F., Queloz, D., et al.\ 2003, The Messenger, 114, 20
    \bibitem[M{\'e}rand(2022)]{pmoired} M{\'e}rand, A.\ 2022, \procspie, 12183, 121831N. doi:10.1117/12.2626700
    \bibitem[Mr{\'o}z et al.(2024)]{Mroz24} Mr{\'o}z, P., Dong, S., M{\'e}rand, A., et al.\ 2024, \apj in press, arXiv:2409.12227
    \bibitem[Onken et al.(2024)]{smssdr4} Onken, C.~A., Wolf, C., Bessell, M.~S., et al.\ 2024, \pasa, 41, e061. doi:10.1017/pasa.2024.53
	\bibitem[Paczy\'nski(1986)]{Paczynski1986} Paczy\'nski, B.\ 1986, \apj, 304, 1. doi:10.1086/164140
    \bibitem[Piskunov \& Valenti(2017)]{Piskunov2017} Piskunov, N. \& Valenti, J.~A.\ 2017, \aap, 597, A16. doi:10.1051/0004-6361/201629124
	\bibitem[Rattenbury \& Mao(2006)]{vltimicrolens3} Rattenbury, N.~J., \& Mao, S.\ 2006, \mnras, 365, 792
    \bibitem[Recio-Blanco et al.(2023)]{RecioBlanco2023} Recio-Blanco, A., de Laverny, P., Palicio, P.~A., et al.\ 2023, \aap, 674, A29. doi:10.1051/0004-6361/202243750
	\bibitem[Refsdal(1966)]{Refsdal1966} Refsdal, s. 1966, \mnras, 134, 315
	\bibitem[Riello et al.(2021)]{Gaiaphotometry} Riello, M., De Angeli, F., Evans, D.~W., et al.\ 2021, \aap, 649, A3. doi:10.1051/0004-6361/202039587
    \bibitem[Rybicki et al.(2022)]{Rybicki2022} Rybicki, K.~A., Wyrzykowski, {\L}., Bachelet, E., et al.\ 2022, \aap, 657, A18. doi:10.1051/0004-6361/202039542
	\bibitem[Schechter et al.(1993)]{Schechter1993} Schechter, P.~L., Mateo, M., \& Saha, A.\ 1993, \pasp, 105, 1342. doi:10.1086/133316
	\bibitem[Shappee et al.(2014)]{Shappee2014} Shappee, B.~J., Prieto, J.~L., Grupe, D., et al.\ 2014, \apj, 788, 48. doi:10.1088/0004-637X/788/1/48
	\bibitem[Sheinis et al.(2015)]{HERMES} Sheinis, A., Anguiano, B., Asplund, M., et al.\ 2015, Journal of Astronomical Telescopes, Instruments, and Systems, 1, 035002. doi:10.1117/1.JATIS.1.3.035002
	\bibitem[Skowron et al.(2011)]{Skowron2011} Skowron, J., Udalski, A., Gould, A., et al.\ 2011, \apj, 738, 87. doi:10.1088/0004-637X/738/1/87
    \bibitem[Skrutskie et al.(2006)]{Skrutskie2006} Skrutskie, M.~F., Cutri, R.~M., Stiening, R., et al.\ 2006, \aj, 131, 1163. doi:10.1086/498708
	\bibitem[Smith et al.(2003)]{Smith2003} Smith, M.~C., Mao, S., \& Paczy{\'n}ski, B.\ 2003, \mnras, 339, 925. doi:10.1046/j.1365-8711.2003.06183.x
    \bibitem[Ting et al.(2019)]{Ting2019} Ting, Y.-S., Conroy, C., Rix, H.-W., et al.\ 2019, \apj, 879, 69. doi:10.3847/1538-4357/ab2331
	\bibitem[Yee et al.(2012)]{Yee2012} Yee, J.~C., Shvartzvald, Y., Gal-Yam, A., et al.\ 2012, \apj, 755, 102. doi:10.1088/0004-637X/755/2/102
	\bibitem[Yoo et al.(2004)]{Yoo2004} Yoo, J., DePoy, D.~L., Gal-Yam, A., et al.\ 2004, \apj, 603, 139. doi:10.1086/381241
	\bibitem[van Cittert(1934)]{vanCittert1934} van Cittert, P.~H.\ 1934, Physica, 1, 201. doi:10.1016/S0031-8914(34)90026-4
    \bibitem[Wyrzykowski et al.(2023)]{Gaiamicrolensing} Wyrzykowski, {\L}., Kruszy{\'n}ska, K., Rybicki, K.~A., et al.\ 2023, \aap, 674, A23. doi:10.1051/0004-6361/202243756
    \bibitem[Zang et al.(2020)]{Zang2020} Zang, W., Dong, S., Gould, A., et al.\ 2020, \apj, 897, 180. doi:10.3847/1538-4357/ab9749
	\bibitem[Zernike(1938)]{Zernike1938} Zernike, F.\ 1938, Physica, 5, 785. doi:10.1016/S0031-8914(38)80203-2
    \bibitem[Zhu et al.(2014)]{Zhu2014} Zhu, W., Penny, M., Mao, S., et al.\ 2014, \apj, 788, 73. doi:10.1088/0004-637X/788/1/73
	
	
\end{thebibliography}
